\documentclass{article}

\usepackage[english]{babel}

\usepackage[letterpaper,top=2cm,bottom=2cm,left=3cm,right=3cm,marginparwidth=1.75cm]{geometry}

\usepackage{natbib}
\usepackage{amsmath}
\usepackage{amssymb}
\usepackage{graphicx}
\usepackage[colorlinks=true, allcolors=blue]{hyperref}
\usepackage{ulem}               
\usepackage{xcolor}
\definecolor{boxes}{HTML}{fff0b3}

\title{Review and prospects of hot exozodiacal dust research for future exo-Earth direct imaging missions}

\begin{document}
\maketitle

\begin{center}
Steve Ertel$^{1,2}$, Tim D.\ Pearce$^{3}$, John H.\ Debes$^{4}$, Virginie C.\ Faramaz$^{1}$, William C.\ Danchi$^{5}$, Ramya M.\ Anche$^{1}$, Denis Defr\`ere$^{6}$, Yasuhiro Hasegawa$^{7}$, Justin Hom$^{1}$, Florian Kirchschlager$^{8}$, Isabel Rebollido$^{9}$, H\'el\`ene Rousseau$^{2,10}$, Jeremy Scott$^{1}$, Karl Stapelfeldt$^{7}$, Thomas A. Stuber$^{1}$
\end{center}

\small
\noindent $^{1}$  Department of Astronomy and Steward Observatory, University of Arizona, 933 N Cherry Ave., Tucson, AZ 85721-0065, USA\\
\noindent $^{2}$  Large Binocular Telescope Observatory, University of Arizona, 933 N Cherry Ave., Tucson, AZ 85721-0065, USA\\
\noindent $^{3}$  Department of Physics, University of Warwick, Gibbet Hill Road, Coventry, CV4 7AL, UK\\
\noindent $^{4}$  Space Telescope Science Institute, 3700 San Martin Drive, Baltimore, MD 21218, USA\\
\noindent $^{5}$  NASA Goddard Space Flight Center, 8800 Greenbelt Road, Greenbelt, MD 20771-2400, USA\\
\noindent $^{6}$  Institute of Astronomy, KU Leuven, Celestijnenlaan 200D, 3001 Leuven, Belgium\\
\noindent $^{7}$  Jet Propulsion Laboratory, California Institute of Technology, Pasadena, CA 91109, USA\\
\noindent $^{8}$  Sterrenkundig Observatorium, Ghent University, Krijgslaan 281 - S9, B-9000 Gent, Belgium\\
\noindent $^{9}$  European Space Agency (ESA), European Space Astronomy Centre (ESAC), Camino Bajo del Castillo s/n, 28692 Villanueva de la Ca\~nada, Madrid, Spain\\
\noindent $^{10}$  AGO Department, University of Li\`ege, All\'ee du 6 ao\^ut, 19C, 4000 Li\`ege 1, Belgium\bigskip\\


\normalsize


\section{Abstract}

Hot exozodiacal dust is dust in the innermost regions of planetary systems, at temperatures around 1000\,K to 2000\,K, and commonly detected by near-infrared interferometry.  The phenomenon is poorly understood and has received renewed attention as a potential risk to a planned future space mission to image potentially habitable exoplanets and characterize their atmospheres (exo-Earth imaging) such as the Habitable Worlds Observatory (HWO).  In this article, we review the current understanding of hot exozodiacal dust and its implications for HWO.  We argue that the observational evidence suggests that the phenomenon is most likely real and indeed caused by hot dust, although conclusive proof in particular of the latter statement is still missing.  Furthermore, we find that there exists as of yet no single model that is able to successfully explain the presence of the dust.  We find that it is plausible and not unlikely that large amounts of hot exozodiacal dust in a system will critically limit the sensitivity of exo-Earth imaging observations around that star.  It is thus crucial to better understood the phenomenon in order to be able to evaluate the actual impact on such a mission, and current and near-future observational opportunities for acquiring the required data exist.  At the same time, hot exozodiacal dust (and warm exozodiacal dust closer to a system's habitable zone) has the potential to provide important context for HWO observations of rocky, HZ planets, constraining the environment in which these planets exist and hence to determine why a detected planet may be capable to sustain life or not.


\section{Introduction}

Study Analysis Group (SAG)~\#23\footnote{\url{https://sites.google.com/view/sag23-exozodiacaldust/home}} ``The Impact of Exozodiacal Dust on Exoplanet Direct Imaging Surveys'' was established as part of NASA's Exoplanet Exploration Program Analysis Group (ExoPAG).  The mandate of SAG~\#23 is to review the available knowledge of dust in planetary systems around mature stars -- in particular in and near the habitable zone (HZ) -- and to study the dust's impact on the search and characterization of rocky, potentially habitable planets and the search for bio-signatures in their atmospheres by a future direct-imaging mission (exo-Earth imaging).

Exozodiacal dust -- or exozodi for short -- is warm dust in and near a star's habitable zone (HZ), with temperatures $\sim$300\,K, and hot dust closer in, with temperatures $\sim$1000\,K to $\sim$2000\,K \citep{kral2017}.  It poses potential obstacles to an exo-Earth imaging mission because it adds noise, confusion, and coronagraphic leakage to these observations \citep{beichman2006a, defrere2010, roberge2012, defrere2012b}.  Understanding the occurrence rate, properties, origin, and dynamics of the dust as a function of other, more accessible system parameters such as stellar properties or the presence of cold dust or giant planets in the system \citep[e.g.,][]{bonsor2012, bonsor2014, bonsor2018, faramaz2017, sezestre2019, rigley2020} can help predict the impact of these noise sources for future mission targets \citep{ertel2020}.  In addition, studying the dust can give important insight into the presence of planets, planetesimals, and comets orbiting a star, and their dynamics and interactions, thereby crucially constraining the architecture and evolution of a planetary system \citep[e.g.,][]{krivov2010}.  This may have profound implications for whether a rocky, HZ planet is Earth-like compared to, e.g., barren or a water world \citep{wyatt2020, kral2018}.

For more than a decade, significant investments from NASA (Keck Interferometer Nuller, KIN, Large Binocular Telescope Interferometer, LBTI, but also, e.g., Spitzer and WISE programs) have led to crucial advances of our understanding of warm exozodiacal dust specifically \citep[e.g.,][]{kennedy2013, beichman2006b, chen2014, millan-gabet2011, mennesson2014, weinberger2015, kennedy2015, defrere2015, ertel2018, ertel2020, defrere2021}.  The conclusion is that the majority of Sun-like stars have relatively low HZ dust levels with 1$\sigma$ and 95\% confidence upper limits of 9 and 27 times the Solar-system level with $\sim$20\% being significantly more dusty \citep{ertel2020}.  While more precise constraints are still required, in particular for the spectroscopic characterization of imaged exo-Earth candidates, it has been established that HZ dust is not a major showstopper for such an endeavor.  During recent efforts to define the scope of ``precursor science'' for a large NASA flagship mission capable of an exo-Earth imaging survey, the importance of also better understanding hot exozodiacal dust was recognized.  Hot exozodi is dust closer to the star than the habitable zone, likely reaching as close as the dust sublimation distance (typically a few to a few 10s of stellar radii depending on stellar luminosity and dust composition, \citealt{kral2017}).  This dust has been detected as a $\sim$1\% near-infrared (nIR) excess around $\sim$20\% of the main-sequence stars surveyed using optical long-baseline interferometry \citep{absil2013, ertel2014, absil2021}.  While the origin and properties of this dust are poorly understood, it is thought that it is supplied by material further out in the planetary system.  This material must then move through the system's HZ, potentially producing emission in the HZ that may limit our ability to detect and characterize rocky planets there.  Furthermore, the hot dust produces extended and likely polarized excess emission around the star \citep{ollmann2023} that may limit the contrast that can be reached by coronagraphic starlight suppression methods.

The presence of hot dust around a significant fraction of stars and/or its persistence over a significant fraction of a star's life time are difficult to explain.  The detections are usually made close to the detection limits of the instruments used and only one technique (optical long-baseline interferometry) has so far been able to reliably detect the dust.  This raises the question whether the detections are real and -- if so -- whether the phenomenon can actually be attributed to the presence of dust.  Many detections have been shown to be repeatable \citep{ertel2016}, and some have been repeated with different instruments and telescopes and at different wavelengths \citep{defrere2011, kirchschlager2020}, supporting the claim that the detections are in fact real and not caused by statistical or systematic errors.  Most recently, observations in $L$~band have shown that the excess emission is indeed consistent with blackbody emission from dust at a temperature of $\gtrsim$1000\,K \citep{kirchschlager2020}.  Furthermore, no scenario other than the presence of dust has been shown to plausibly explain the excess emission so far.  This all strongly supports the conclusion that the excess emission is in fact real and most likely caused by the presence of hot exozodiacal dust.

In this paper we review the current knowledge of hot exozodiacal dust and open questions, the current state of research in the field, and future prospects of improving our understanding of the dust in support of a future, large exo-Earth imaging mission.  We first define hot exozodiacal dust in Sect.~\ref{sec:definition} and describe the distinction from other dust species commonly discussed in the context of mature exoplanetary systems.  In Sect.~\ref{sec:observations} we describe the observational methods used to study hot exozodi and summarize the observational results.  The theoretical challenges and efforts to understand hot exozodi are summarized in Sect.~\ref{sec:theory}.  In Sect.~\ref{sec:exo-earth} we predict the impact of hot exozodi on a future exo-Earth imaging mission.  Future observational prospects to further our understanding of hot exozodi are described in Sect.~\ref{sec:prospects}.  We summarize our findings in Sect.~\ref{sec:summary}.


\section{Definition of hot exozodiacal dust and distinction from other dust species around mature stars}
\label{sec:definition}
\begin{figure}
  \centering
  \includegraphics[width=\textwidth]{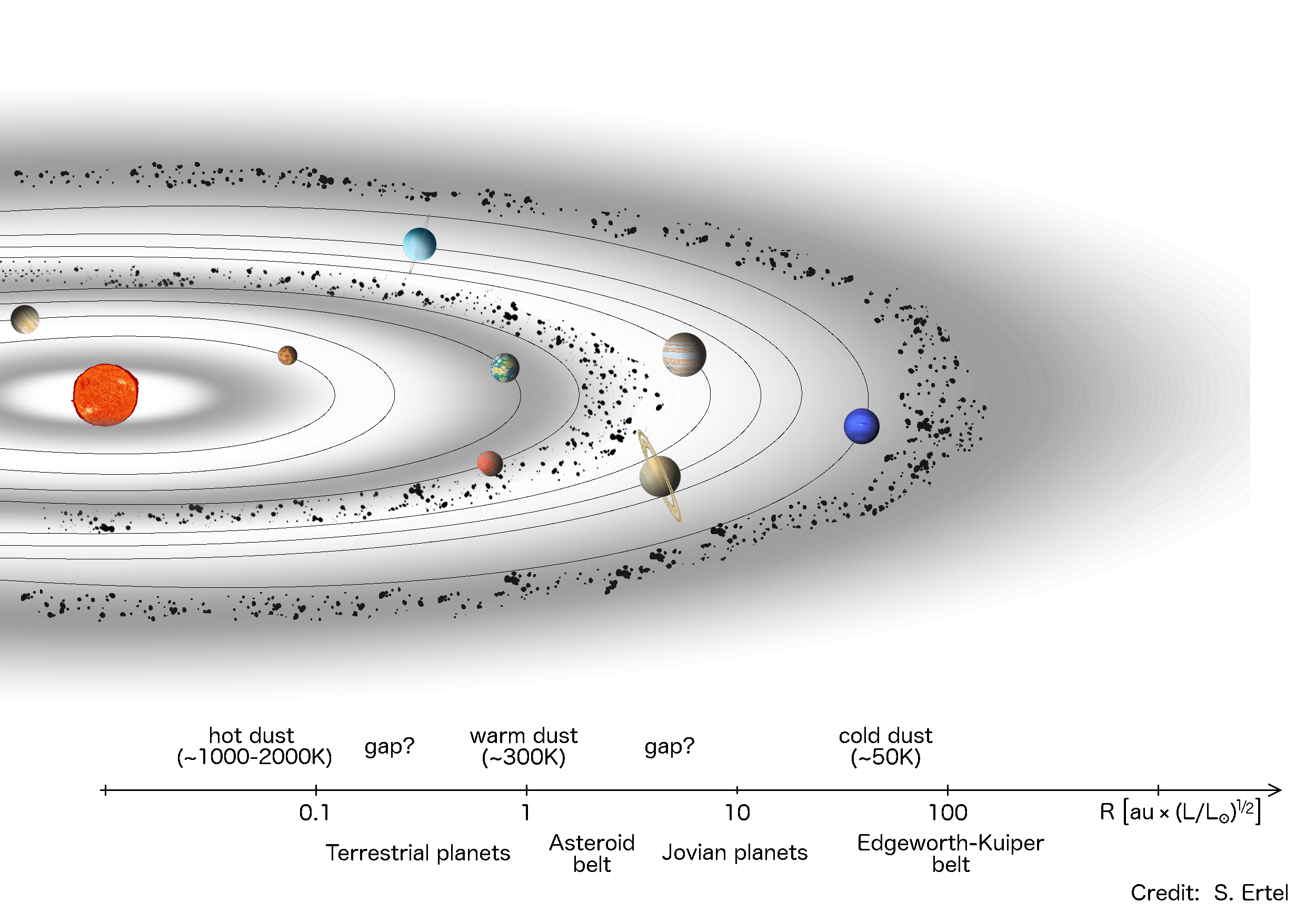}
  \caption{Sketch illustrating the distinction between hot and warm (HZ) exozodiacal dust and cold dust, as well as their location relative to other components of planetary systems.  The Solar system is used as an approximate template, though the exact amount of hot dust in the Solar system is poorly known.  Broad, gray rings represent dusty regions, while dotted, black rings represent belts of minor bodies (asteroids, Kuiper-belt objects).  Possible planets and their orbits (thin black lines) similar to those in the Solar system are also shown.}
  \label{fig_zodi_sketch}
\end{figure}

\begin{figure}
  \centering
  \includegraphics[width=0.7\textwidth]{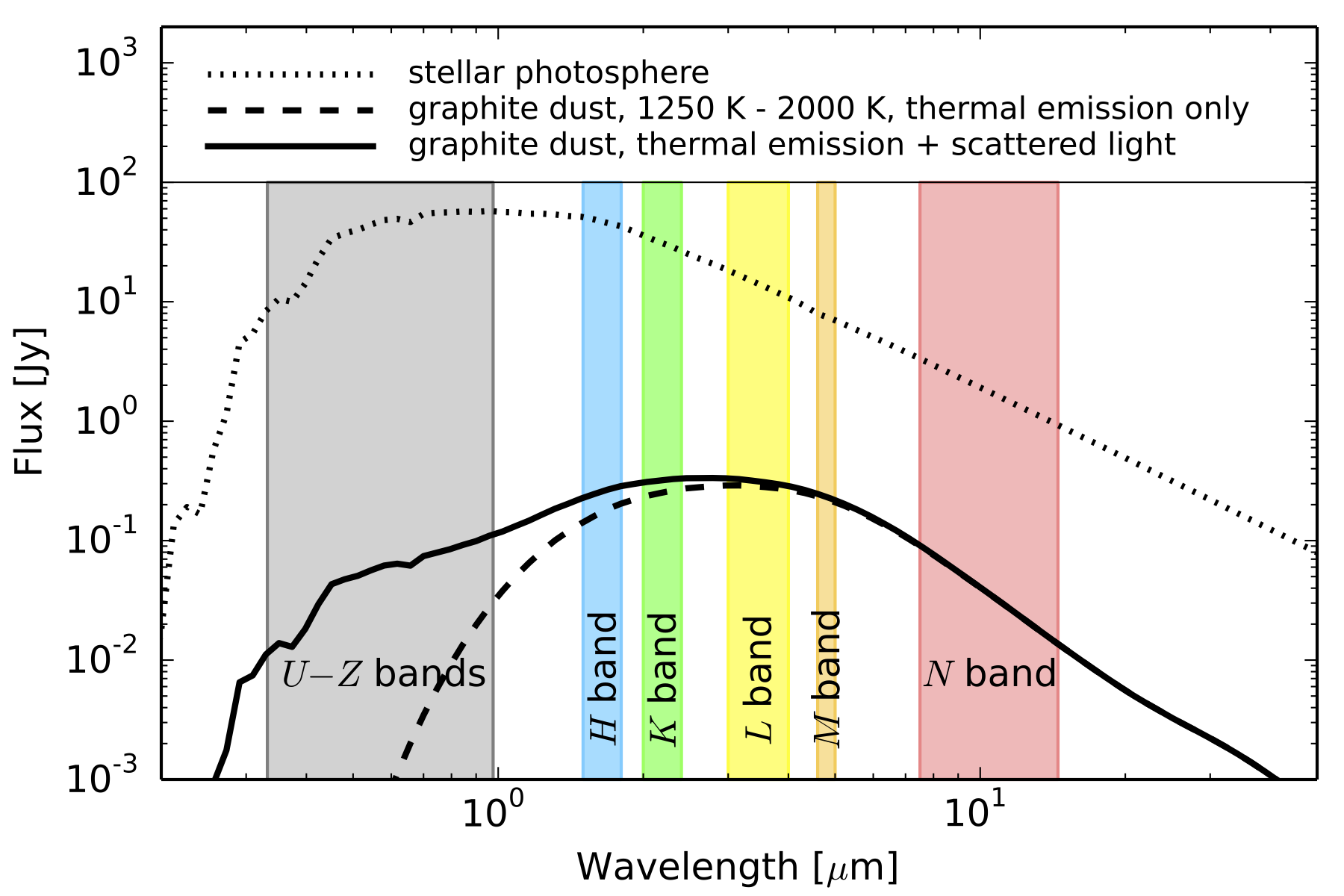}
  \caption{Contribution of scattered light and thermal emission to the light from hot exozodiacal dust.  A plausible toy model around a Sun-like star at 10\,pc has been used, assuming carbon grains ranging from $0.5\,\mu$m to $1.0\,\mu$m in size and from the sublimation radius (assumed to be located where the dust grains reach 2000\,K) to 0.05\,au from the star, where the coolest grains have a temperature of 1250\,K.  The dust mass has been scaled to produce 1\% excess in the $K$~band.  Simulations were done using DDS \citep{wolf2005}.  A similar simulation has been presented by \citet{diFolco2007} for the specific hot-dust detection around the star $\tau$\,Ceti.}
  \label{fig_therm_sca}
\end{figure}

Fig.~\ref{fig_zodi_sketch} shows the different types of dust known to be potentially present in planetary systems and their location relative to other components of planetary systems.  Exozodiacal dust, i.e., dust in the inner regions of planetary systems, can be observationally separated into two classes, warm dust in and near the HZ of its host star (temperatures of $\sim$300\,K) and hot dust closer in (temperatures of $\sim$1000\,K to $\sim$2000\,K).  The location and temperature are the known, distinguishing properties of the two classes.  Warm, HZ dust predominantly emits in the mid infrared (mIR) and is commonly detected using $N$-band (wavelength range 8-13\,$\mu$m) nulling interferometry with moderately high angular resolution \citep[e.g.,][]{mennesson2014, ertel2020}, as a separation of typically 1\,au from a star at a typical distance of 10\,pc corresponds to an angular separation of 0.1\,arcsec.  Hot dust closer in emits predominantly in the nIR and its detection requires the use of high-angular resolution (0.1\,au at 10\,pc corresponds to 0.01\,arcsec), commonly achieved by optical long-baseline interferometry in the $H$~and $K$~bands (wavelength range 1.6-2.2\,$\mu$m, \citealt{absil2006, akeson2009, defrere2011, absil2013, ertel2014, ertel2016, nunez2017, absil2021}), and recently in the $L$~band (wavelength range 3-4\,$\mu$m, \citealt{kirchschlager2020}).  It is plausible that scattered light also contributes to the detected excess, in particular in the $H$~band \citep{ertel2014}, as demonstrated in Fig.~\ref{fig_therm_sca}.\bigskip

\noindent\colorbox{boxes}{\fbox{\begin{minipage}{0.97\textwidth}
\textbf{Definition.}~~Hot exozodiacal dust is dust in the innermost regions around mature stars at temperatures around 1000\,K to 2000\,K.  Its defining observational feature is its high temperature that results in detectable thermal emission in the $H$~and/or $K$~band (wavelengths around 1.6\,$\mu$m and 2.2\,$\mu$m).
\end{minipage}}}\bigskip

No clear observational connection between the hot dust and HZ dust, such as a statistical correlation or anti-correlation between the presence of the two species, has been made so far \citep{kirchschlager2017, kirchschlager2018, ertel2020, absil2021}.  Neither detection method is currently capable of detecting the other dust species, respectively:  warm dust does not sufficiently emit in the nIR to be detected with current long-baseline interferometry in that wavelength range and available nulling interferometric data at mIR wavelengths do not have sufficient angular resolution or sensitivity to detect the hot dust (see Sect.~\ref{sec:observations} for details).

Hot exozodiacal dust is also distinct from dust further out in the system and at even lower temperatures, commonly attributed to the presence of a belt of planetesimals analog to the Solar system's asteroid belt or Edgeworth-Kuiper belt and typically detected as an excess above the stellar photospheric emission in the mIR \citep[e.g.,][]{beichman2006b, carpenter2009, lawler2009, chen2014, morales2012, patel2014, cotten2016} to far-infrared \citep[fIR, e.g.,][]{chen2005, beichman2006a, bryden2006, hillenbrand2008, booth2013, eiroa2013, gaspar2013, montesinos2016, sibthorpe2018}.  A correlation between the presence of hot dust and cold dust in the observed systems was found by \citet{nunez2017} and \citet{absil2021}, potentially suggesting an origin of the hot dust material in the outer regions of a planetary system.  This connection is further supported by the short-lived nature of the hot dust as material on very short orbits is prone to rapid collision, sublimation, accretion, and radiation-pressure blow-out, which suggest the need for a reservoir \citep[e.g.,][]{backman1993, wyatt2007} to replenish the hot dust \citep[e.g.,][]{vanlieshout2014, rieke2016, kimura2020, pearce2020, pearce2022}.

Finally, we distinguish exozodiacal dust from the dust detected in extreme debris disks as described, e.g., by \citet{balog2009, meng2012, su2019}, and the dust observed as variable absorption towards main-sequence stars \citep{boyajian2016, strom2020, marshall2023b}.  While the dust in these systems may also be warm or hot and located close to the star, it is thought to be more abundant and its presence is typically attributed to a rare, transient event such as a planetary collision or an exocometary transit.  In contrast, the warm and hot exozodiacal dust considered here is comparably more tenuous and found around $\sim$20\% of the observed stars, suggesting it is either long-lived or supplied by a mechanism that is fairly common over the lifetime of a significant fraction of stars.


\section{Observing hot exozodiacal dust}
\label{sec:observations}
The defining observational property of hot exozodiacal dust is its (most likely) thermal emission in the nIR ($H$~or $K$~bands, 1.6\,$\mu$m to 2.2\,$\mu$m).  This means the dust must be very hot and hence very close to its host star, well within 1\,au for most stars.  The emission region has a small angular size (0.1\,au at a typical distance of 10\,pc corresponds to an angular scale of 0.01\,arcsec) and the hot emission is observed at wavelengths where it is outshone by the host star.  These properties are distinctly different from those of the emission from cold, Kuiper-belt-like debris disks, which peaks at fIR wavelengths where it is easily distinguished from the host star's emission.  Typical flux ratio between detected hot exozodi and its host star is $\sim$1\% in the $H$~or $K$~band.  This makes the dust the most luminous component of the planetary system after the star itself; yet photometric detection of this excess remains mostly elusive due to calibration uncertainties and the precision at which the stellar photospheric flux can be predicted.  Thus, detecting hot exozodi typically requires observationally disentangling the dust emission from the star's light.  In this section we describe the observational methods commonly used to detect hot exozodiacal dust and the basic observational results.

\subsection{Detection methods}
\label{sec:detection_methods}

\begin{table}[t]
\centering
\begin{tabular}{l l l l}
\hline
  Instrument & Detection method & Wavelength range & Pertinent references \\
\hline
  CHARA/FLUOR  & long-baseline interferometry & $K$~band & 1, 2, 3, 4 \\
  VLTI/VINCI   & long-baseline interferometry & $K$~band & 5 \\
  IOTA/IONIC   & long-baseline interferometry & $H$~band & 6 \\
  PFN          & nulling interferometry & $K$~band & 7 \\
  KIN          & nulling interferometry & $N$~band & 8, 9 \\
  VLTI/PIONIER & long-baseline interferometry & $H$~band & 10, 11, 12, 13 \\
  AAT/HIPPI    & integral polarimetry & visible & 14 \\
  CHARA/JouFLU & long-baseline interferometry & $K$~band & 15 \\
  LBTI         & nulling interferometry & $N$~band & 16, 17, 18, 19 \\
  VLTI/MATISSE & long-baseline interferometry & $L$~band ($L$, $M$, $N$~bands)$^\dag$ & 20 \\
  JWST/MIRI MRS & spectroscopy & 5-7.5\,$\mu$m (5-28\,$\mu$m$^\dag$) & 21 \\
\hline
\end{tabular}
\caption{List of instruments with major, published contributions to the study of hot exozodiacal dust as described in Sect.~\ref{sec:detection_methods}, in rough chronological order.\\
$^\dag$ The wavelength range in which the relevant results were obtained is listed first, the broader operating range of the instrument is listed in parentheses.\\
References.  (1)~\citet{absil2006}, (2)~\citet{diFolco2007}, (3)~\citet{akeson2009}, (4)~\citet{absil2013}, (5)~\citet{absil2009}, (6)~\citealt{defrere2011}, (7)~\citep{mennesson2011}, (8)~\citet{millan-gabet2011}, (9)~\citet{mennesson2014}, (10)~\citet{defrere2012a}, (11)~\citet{ertel2014}, (12)~\citet{ertel2016}, (13)~\citet{absil2021}, (14)~\citet{marshall2016}, (15)~\citet{nunez2017}, (16)~\citet{defrere2015}, (17)~\citet{ertel2018}, (18)~\citet{ertel2020}, (19)~\citet{defrere2021}, (20)~\citet{kirchschlager2020}, (21)~\citet{worthen2024}.}
\label{table:exozodi_instruments}
\end{table}

In the following, we describe the main methods that have been used to detect hot exozodiacal dust or to put strong observational constraints on the dust properties from meaningful non-detections.  A summary list of instruments with major, published contributions to the study of hot exozodiacal dust is presented in Table~\ref{table:exozodi_instruments}.  Current and near-future capabilities that have not yet been exploited but could provide significant observational advances in the field will be discussed in Sect.~\ref{sec:prospects}.

\subsubsection{Optical long baseline interferometry}

The first detections of hot exozodiacal dust have been made using optical long-baseline interferometry (OLBI, \citealt{absil2006}), and it remains by far the most successful method for observing the phenomenon \citep[e.g.,][]{absil2013, ertel2014, kirchschlager2020, absil2021}.  While reaching the required precision to detect the typical 1\% dust-to-star flux ratio with these observations is still challenging, it has been demonstrated with a few high-precision instruments.  OLBI has thus so far been the only method employed to successfully and consistently detect hot exozodiacal dust around a large sample of targets.

\begin{figure}
  \centering
  \includegraphics[width=0.7\textwidth]{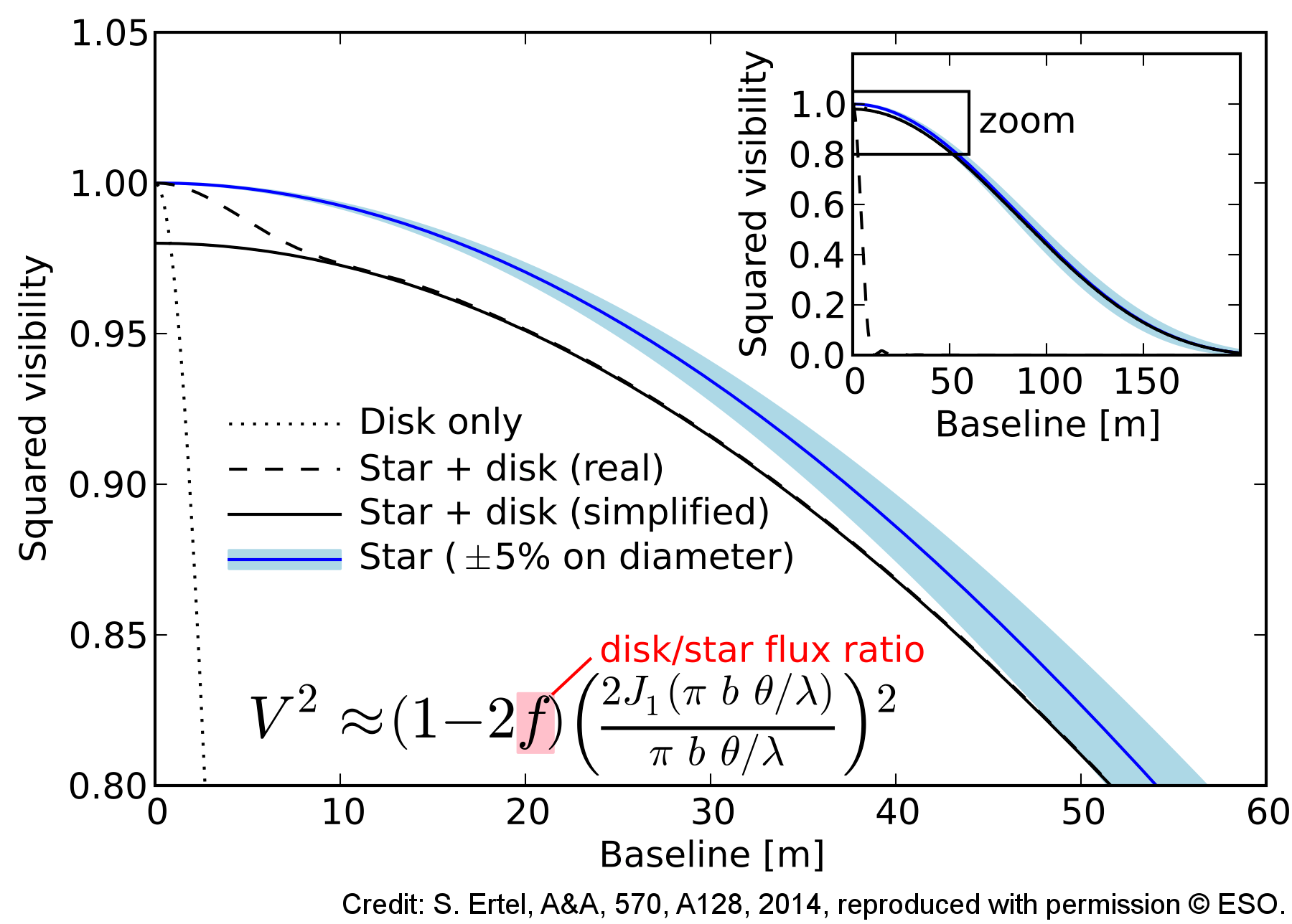}
  \caption{Detection principle of hot exozodiacal dust using optical long-baseline interferometry \citep[taken from][]{ertel2014}.  At intermediate baselines ($\sim$10-30\,m in the $H$~band) the star remains largely unresolved, mitigating uncertainties on its diameter when predicting its visibilities.  A dust disk (or cloud) is well resolved and produces a visibility drop approximated as twice the disk-to-star flux ratio.}
  \label{fig_nir_detection_method}
\end{figure}

Detecting exozodiacal dust with OLBI makes use of the difference in visibility between extended (resolved) and compact (unresolved) emission regions (Fig.~\ref{fig_nir_detection_method}).  At nIR wavelengths and baselines of a few 10\,m, a main-sequence star remains largely unresolved.  Its visibility is thus close to one and the impact of its uncertain diameter on the prediction of its visibility is small.  On the other hand, extended emission from hot exozodi is well resolved, adding incoherent emission that yields a very low visibility and the impact of the exact shape of the emission on the measurement is small.  In the simplified case of fully resolved dust emission (i.e., the emission homogeneously fills the interferometric field of view) around an unresolved star, this results in a visibility deficit that is twice the dust-to-star flux ratio \citep{diFolco2007}.  It is important to understand that this measurement of the visibility deficit results in a relative flux ratio between circumstellar emission and star and is hence not affected by uncertainties on predicting or measuring the stellar brightness.  In practice, the stellar diameter and its uncertainty are taken into account when determining the observed visibility deficit.  \citet{absil2021} have challenged the assumption of fully resolved dust emission and found that, while leading in most cases to an underestimation of the disk-to-star flux ratio, it has no significant impact on the conclusion of their and previous work.  \citet{kirchschlager2018} modeled the observational signatures of specific spatial dust distributions and concluded that constraining the radial location of the dust may be in reach of near-future observations.  Recent results by Priolet et al. (in prep.) and Stuber et al. (in prep.) suggest the first signs of spatial information about the dust geometry have been found in data obtained with the Multi-AperTure mid-Infrared SpectroScopic Experiment (MATISSE; \citealt{lopez2012, lopez2022}) on the Very Large Telescope Interferometer (VLTI).  Ollmann et al. (forthcoming) have introduced a self-consistent modeling approach of the visibilities directly.  That will allow for circumventing the assumption of fully resolved emission once suitably constraining data are available.

Detecting hot dust at the typical 1\% level of excess (2\% visibility deficit) requires individual visibility measurements to the precision of at least 1\% if multiple measurements can be combined to boost the precision of the final result (i.e., if measurements are sufficiently independent and unbiased).  This puts strong requirements on the instrumental precision, the precision of the absolute calibration of the data, and the suppression and/or characterization of systematic errors that would limit the ability to combine multiple observations for higher cumulative precision.  The first instrument routinely used for detecting hot exozodiacal dust with sufficient precision was the Fiber-Linked Unit for Optical Recombination (FLUOR, \citealt{coudeduforesto2003}) beam combiner on the Center for High-Angular Resolution Astronomy (CHARA) array \citep{absil2006, absil2013}.  FLUOR used two telescopes, thus producing a single visibility measurement per observation which limited its efficiency.  Nonetheless, FLUOR produced the first survey for hot exozodiacal dust from 2006 to 2013.  While other instruments (VLTI/VINCI, \citealt{absil2009} and IOTA/IONIC, \citealt{defrere2011}, both now decommissioned) have made small contributions, a breakthrough was achieved with the Precision Integrated Optics Near Infrared ExpeRiment (PIONIER, \citealt{lebouquin2011}) beam combiner on the VLTI.  PIONIER uses four telescopes, simultaneously producing six visibility measurements.  This improved observing efficiency has resulted in the largest exozodi survey yet with almost 100 stars observed in 2012 \citep{ertel2014} alone and more observations taken mostly between 2013 and 2016 \citep{ertel2016, absil2021}.  Both FLUOR and PIONIER have routinely reached a precision of $\sim$1\% on individual visibility measurements.  FLUOR underwent an upgrade (JouFLU project), but has since not been able to deliver the same precision and is currently unavailable due to technical problems.  While it is possible that past performance can be achieved again in the future, this appears unlikely without a major effort.

Measuring the visibility deficit requires precise calibration.  A suitable calibration method has been formalized and studied in detail for PIONIER by \citet{ertel2014}.  A large, unbiased sample of stars has been observed and the data have been empirically evaluated for correlations and biases.  The authors conclude that multiple measurements of a science target (typically of the order of 10 to 20, even if taken simultaneously on different baselines) can be combined yielding a final precision per target of the order of 0.2\%.  This is achieved by interleaving observations of a science target (SCI) with observations of reference stars (CAL) in a rapid CAL1--SCI--CAL2--SCI--[...] sequence with typically 10-15\,minutes per star.  Giant stars are used as calibrators due to the low probability of them having companions that would be detectable at IR wavelengths and interfere with the calibration.  Several different calibration stars are used per science target to further minimize the risk of bad calibrators (binaries, stars with significant circumstellar emission) and the impact of uncertain calibrator diameters.

One challenge of any interferometric observations (including nulling interferometry, see Sect.~\ref{sect_nulling}) is spatial filtering.  OLBI has an inner working angle and interferometric field of view, so that dust too close or too far from the star will not be detected.  Nulling interferometry in addition has a transmission pattern that filters out signal even across the field of view and outside the inner working angle.  This needs to be taken into account when preparing observations (selecting suitable baseline lengths for a given system and observing wavelength) and when analyzing the data (correcting for any potential loss of signal).  \citet{absil2021} studied the effect of the limited inner working angle for PIONIER observations and found that in their data the baseline length was generally well-chosen.  Yet, a small bias in their detection statistics had to be corrected if the dust was assumed to be located near the closest plausible sublimation radius.  In the future when more sensitive data become available, baseline orientation with respect to the orientation of a brightness distribution on sky that is potentially not azimuthally symmetric (e.g., edge-on disk, dust clumps) will have to be taken into account.

\subsubsection{Precision spectroscopy}
Precision spectroscopy over a wide wavelength range from the visible to the mIR can, in principle, detect the thermal emission from the dust by simultaneously characterizing both the host star's photosphere (hence resulting in a strong constraint on its brightness and spectral slope across the wavelength range used) and detecting subtle deviations of the predicted spectral slope of the flux due to a dust excess.  The advantage of this method is that it measures the integrated flux (no spatial filtering) and the spectral shape of the emission.  It can thus help alleviate some of the challenges caused by the spatial filtering of interferometry and provide a densely-sampled spectral energy distribution of the dust, including the potential detection of spectral features indicative of different dust species and properties.  It remains to be demonstrated that the required precision to detect a large sample of hot exozodi systems can be reached consistently and that the data are free of significant systematics and calibration uncertainties.  A survey using IRTF/SpecX has been executed and data analysis is in progress (C.~Lisse, personal communication).

\citet{worthen2024} report the detection of hot excess emission in JWST/MIRI Medium Resolution Spectrograph (MRS) observations of $\beta$\,Pic in the 5-7.5\,$\mu$m wavelength range and attribute this to the presence of the previously-detected hot dust \citep{defrere2012a}.  More such observations may be forthcoming (I.~Rebollido, personal communication), potentially opening up a new avenue for characterizing the spectral shape of the dust emission for some of the brightest excesses.  A detailed characterization of systematic and calibration uncertainties will have to be performed to evaluate the suitability of this method for detecting and characterizing a large sample of exozodi systems.

\subsubsection{Polarimetry}
\label{sect_integral_polarimetry}
Another method to separate the stellar and dust emission is precision polarimetry.  The integrated light from main sequence stars is generally largely unpolarized, while scattered light from the dust may be polarized depending on its spatial distribution and scattering angle.  Challenges are the small polarization expected from the dust's scattered light on top of the bright star and the contribution from interstellar polarization on the line-of-sight.  This method has been used by \citet{marshall2016} using the High-Precision Polarimetric Instrument (HIPPI) on the Anglo-Australian Telescope (AAT) at visible-light wavelengths.  No signatures of exozodiacal dust were detected, yet important constraints on the origin of the emission and the distribution of the dust were derived (Sect.~\ref{sect:obs_thermal_dust}).  

Previously, unresolved polarimetric observations of the stars with IR excess/debris disks have been carried out by \cite{eritsyan2002, tamura2005, wiktorowicz2010, vandeportal2019}.  \cite{chavero2006optical} compiled a list of polarimetry data of 125 stars with IR excess at 60$\mu$m and 100$\mu$m (38 from their observations and 87 from \citealt{bhatt2000polarization, heiles20009286, oudmaijer2001export}) to analyze the polarization excess.  Among 71 stars within distance of 300\,pc, 37 (52\%) stars show results consistent with no intrinsic polarization and 34 (48\%) stars have a polarization excess over the interstellar components indicating the presence of intrinsic polarization. Further, 20\% of these 34 polarized stars have a polarization $>0.1\%$.  \citet{marshall2023} also detected polarimetric signals from stars with bright debris disks. The differences between the works by \citet{marshall2016} compared to \cite{chavero2006optical} and \citet{marshall2023} suggest that the detected polarimetric signal originates from the outer disk and not from a potential hot-dust component.

The non-detections from \citet{marshall2016} and significant upper limits on intrinsic polarization of $\sim70$\,ppm have several consequences.  First, they seem to rule out that the vast majority of the $\sim1$\% excesses seen in $H$~band are produced by scattered light.  If scattered light was the main source of the $H$~band detections, this would be expected to result in a similar $\sim1$\% excess at visible wavelengths assuming gray-scattering dust grains.  That would mean that the intrinsic, integral polarization of the light scattered by the dust could be no more than $\sim0.7$\% \footnote{The conversion from the 70\,ppm upper limit on the polarization of the \emph{total light} received from the system (direct + scattered star light) to the upper limit on the polarization of the \emph{scattered light} depends on the assumed fraction of scattered-light.  Assuming 1\% scattered light, then its polarization has to be 0.7\% to produce 70\,ppm polarization of the total light.}.  Furthermore, as seen in Fig.~\ref{fig_therm_sca}, a significant amount of scattered-light from the dust at visible wavelengths is expected, even if the majority of the nIR excess is caused by thermal emission.  Assuming 0.1\% scattered light would limit the total, integral polarization of the light scattered by the dust grains to $\sim7$\%.

Those numbers have to be considered order-of-magnitude estimates due to the significant uncertainty of our knowledge of the scattering properties of the dust.  Moreover, the integral polarization of the scattered light from the system could be much lower than the actual polarization of the light scattered by individual dust grains due to geometric effects.  The dust could be distributed in a disk that is seen face-on (unlikely the case for all observed targets) or in a spherical cloud.  In the absence of sufficient clumpiness in the dust distribution, the polarization vectors would then mostly cancel out in spatially unresolved observations.  This could be overcome by spatially resolving the dust distribution.  Alternatively, it is conceivable that the dust properties are such that it does not strongly polarize scattered light, which seems consistent with the presence of sub-micron sized dust grains \citep{marshall2016} commonly found from other studies of hot exozodi (Sect.~\ref{subsec: lackOfMIRDetections}).  In conclusion, these data add an important piece of the puzzle, though their stand-alone interpretation remains unclear.

\subsubsection{Nulling interferometry}
\label{sect_nulling}
Nulling interferometry is commonly known as a method to detect warm (HZ) exozodiacal dust in the $N$~band (Sect.~\ref{sec:definition}).  The
two main instruments used for these observations are the Keck Interferometer Nuller (KIN, \citealt{millan-gabet2011, mennesson2014}, now decommissioned) and the Large Binocular Telescope Interferometer (LBTI, \citealt{defrere2015, ertel2018, ertel2020, defrere2021}).  This method should in principle also be capable of detecting the thermal emission of the hot dust at these wavelengths depending on the spectral slope of the emission.  The requirement is to resolve the same spatial scales as nIR OLBI at an angular resolution proportional to $\lambda/B$, where $\lambda$ is the wavelength and $B$ is the interferometric baseline.  Given $\lambda=1.6\,\mu$m or $\lambda=2.2\,\mu$m and typical baselines $\gtrsim10$\,m for OLBI and $\lambda=8\,\mu$m to $\lambda=13\,\mu$m for $N$-band nulling interferometry, the nulling baseline thus needs to be at least of the order of 35\,m.  This was possible with KIN ($B=85$\,m or less due to sky projection), but excludes LBTI with $B=14.4\,$m. Thus, while more sensitive to emission it can resolve, the LBTI is typically not capable to resolve the hot dust if it is located close to the inner working angle of the OLBI observations used to detect it.  Photometry and spectroscopy could be employed and compared to nulling observations to further constrain the location of the dust and reach inside the inner working angle of the interferometric observations, but only for very strong excesses that are likely not representative of typical systems \citep{lebreton2016}.

There are several projects applying nulling interferometry at shorter wavelengths.  In principle, the higher contrast compared to constructive OLBI reachable by suppressing the bright starlight provides an advantage.  Most noteworthy in the field of exozodiacal dust is the Palomar Fiber Nuller (PFN, \citealt{martin2008}) which has been used to constrain the location of the hot dust around Vega \citep{mennesson2011}.  The GLINT integrated-photonic nulling interferometer \citep{norris2020} on the single-aperture SUBARU telescope, operating in the $H$~band, is another potential avenue for further characterizing hot exozodiacal dust, although its utilization and performance for such observations remains to be studied and demonstrated.  In the near future, Asgard/NOTT will become available at the VLTI and will perform nulling observations at $L$~band \citep{defrere2022, laugier2023, martinod2023}.

\subsubsection{Aperture masking interferometry}
Aperture masking is another interferometric method that could possibly be used to detect exozodiacal dust.  On current large telescopes, this can provide maximum baselines of 6-8\,m, which can be complementary to the baselines provided by OLBI \citep{kirchschlager2018} and possibly capable of constraining the overall shape of the spatial dust distribution (e.g., spherical vs.\ disk-like) due to larger spatial scales measured by the dense, instantaneous coverage of the interferometric u-v-plane with relatively short baselines.  However, ground-based aperture masking commonly suffers from insufficient visibility calibration and the commonly used closure phases from these observations are not able to well-detect any potentially smooth and point-symmetric dust distributions.  Aperture masking from space with JWST may be able to alleviate this challenge (see Sect.~\ref{sec:jwst_ami}).

\subsection{Observational results}
In this section, we summarize the observational results.  Once one has established that the excesses are real and intrinsic to the targets, and that it is likely caused by thermal emission from circumstellar dust, the strong near-infrared emission and lack of detectable mid-infrared excess lead one to conclude that the dust must be very hot and composed of small, sub-micron sized grains.  This was first established by \citet{absil2006} and has since been confirmed by a multitude of studies.  The detection rate of excesses from hot dust is around $\sim$20\% of observed main-sequence stars, larger for the youngest stars but otherwise showing no significant clear dependence.  A mild correlation with the presence of cold dust in the outer regions of the systems is found.  Variability has been detected in one system and tentatively detected in a few more, suggesting variation of the amount and/or temperature (correlated with radial location and/or dust grain size) on time scales of one year or less.  In the following, we describe these results in more detail.

\subsubsection{Excesses are real and intrinsic to the targets}
The challenges of observing hot exozodi described in the previous section result in a paucity of observational constraints.  Relatively large exozodi surveys using nIR OLBI have provided the strongest observational results so far, but most detections are achieved close to the detection limits around an excess of $\sim$1\% at a limited significance around 3-5\,$\sigma$.

\begin{figure}
  \centering
  \includegraphics[width=0.7\textwidth]{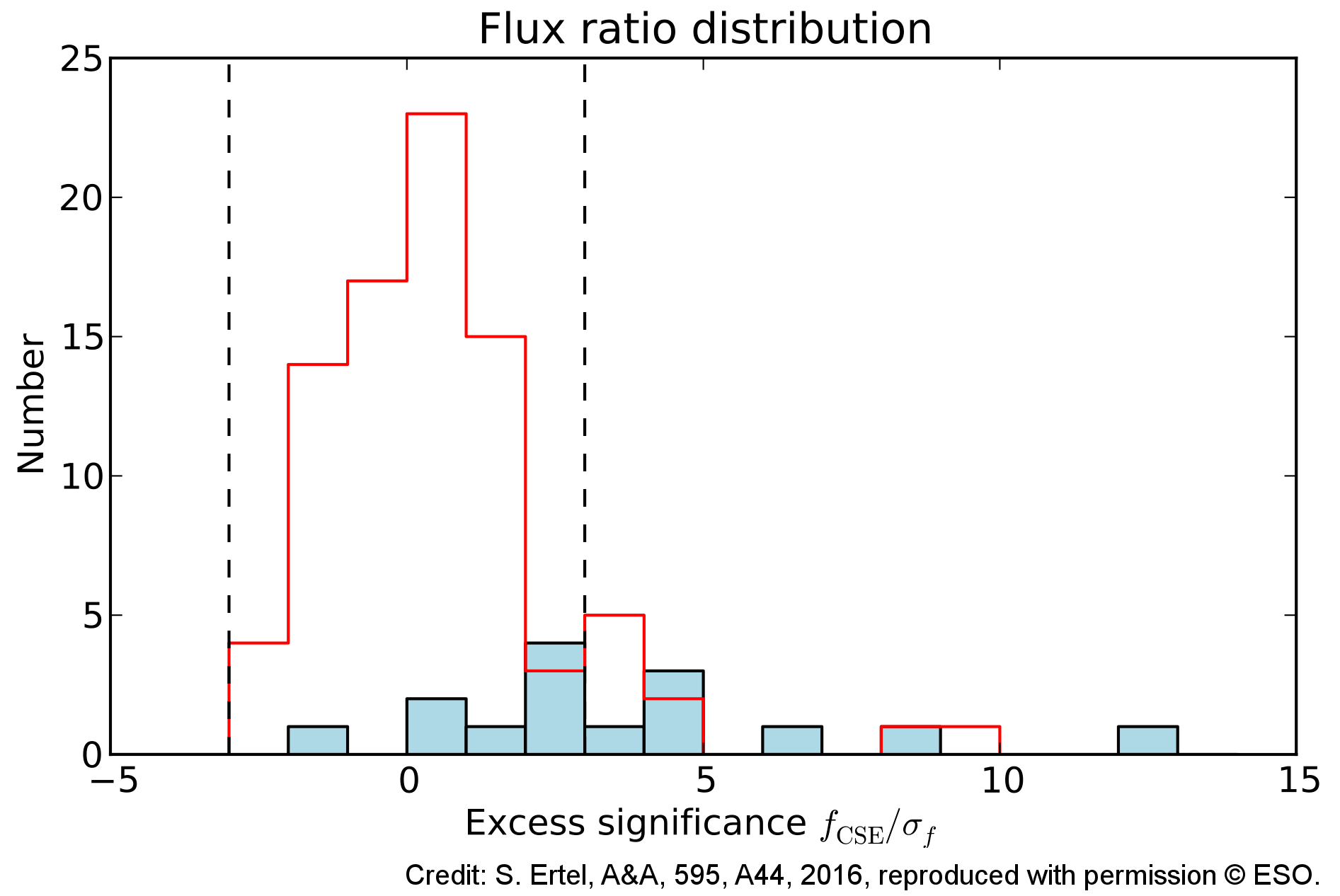}
  \caption{Demonstration of repeatability of the nIR-interferometric detections with PIONIER \citep{ertel2016}.  The red line shows the excess significance (excess measurements divided by its error) for the full, blind survey from \citet{ertel2014}.  The blue histogram shows the measurements from repeated observations of original detections.  While not all excesses are re-detected, which was attributed to the limited signal-to-noise ratio of the original and new detections and potential variability, the excess significance of the repeated observations is significantly skewed to positive excesses. \citet{ertel2016} used a two-sample Anderson-Darling test to demonstrate that the probability that the two histograms are drawn from the same distribution is $5.5\times10^{-5}$, i.e., the distribution of the excess significance from the follow-up observations is not consistent with that from the original, blind survey.}
  \label{fig_repeatability}
\end{figure}

\citet{absil2013}, \citet{ertel2014}, \citet{nunez2017}, and \citet{absil2021} have executed such surveys with FLUOR and PIONIER.  These two instruments have observed 133 stars (PIONIER) and 40 stars (FLUOR) to sufficient sensitivity and detected a total of 33 hot exozodi systems (total detection rate 19$\pm$3\%).  \citet{ertel2016} have statistically demonstrated through repeated observations that the detected signal is intrinsic to the target stars rather than a systematic instrumental error, despite the challenges related to the low S/N detections and the likely variability of the excess seen in at least one source (Fig.~\ref{fig_repeatability}).  Only four stars have been observed by both FLUOR and PIONIER; only one of them, $\lambda$\,Gem, has a FLUOR detection and none have a PIONIER detection.  Individual, known exozodi systems were observed with some success \citep{absil2009, defrere2011} with VINCI on the VLTI and IONIC on the Infrared-Optical Telescope Array (IOTA).  Preliminary results from SpeX on NASA's Infrared Telescope Facility (IRTF) seem to re-detect the interferometric excesses (C. Lisse, personal communication).  We thus conclude that the detected excess emission is likely real and intrinsic to the target stars.

\subsubsection{Excesses stem predominantly from hot thermal emission from dust grains}
\label{sect:obs_thermal_dust}
\begin{figure}
  \centering
  \includegraphics[width=0.7\textwidth]{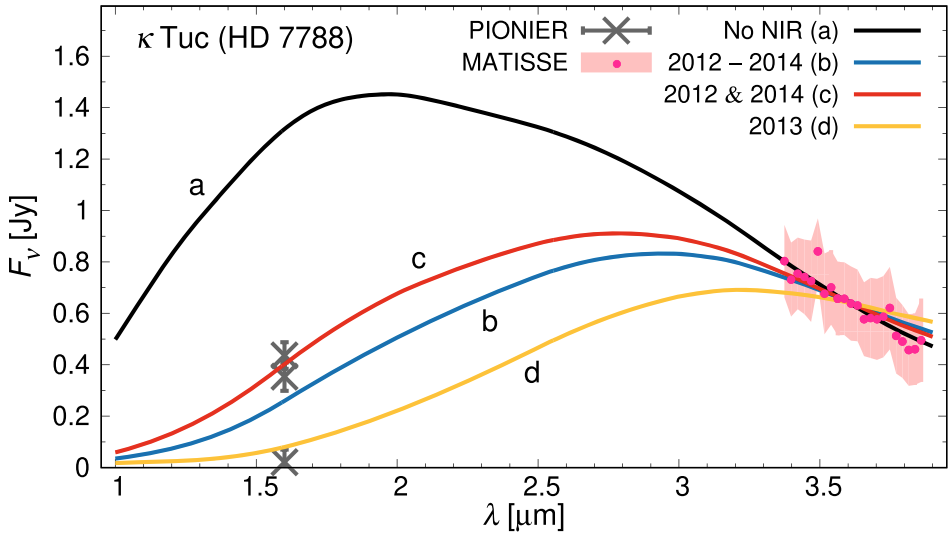}
  \caption{Near-infrared excess measurements for $\kappa$\,Tuc from \citet{kirchschlager2020}.  VLTI/PIONIER measurements at $1.6\,\mu$m, taken about one year apart, yield three different fluxes, which was attributed to variability (gray crosses; \citealt{ertel2016}). VLTI/MATISSE data, taken several years later, yield a spectral slope at longer wavelengths (pink points).  Since the $1.6\,\mu$m flux level is unknown at the time of the MATISSE observation, the hot-exozodi SED could have many different shapes (lines a-d), though all fits are consistent with hot thermal emission from dust grains.  Image reproduced from \citet{kirchschlager2020}.}
  \label{fig: kirchschlager2020Fig3}
\end{figure}

The detection rate increases from $H$~to $K$~band from $17^{+4}_{-3}\%$ to $28^{+8}_{-6}\%$.  While this trend has to be considered marginal given the large statistical uncertainties, it suggests that the dust-to-star flux ratio is more favorable at longer wavelengths, i.e., the emission is redder than the star as would be expected from thermal dust emission.  This has been confirmed by \citet{kirchschlager2020} who detected the hot dust around $\kappa$\,Tuc with MATISSE in the $L$~band and show that the spectral energy distribution of the emission is well fit by dust grains with temperatures between 900\,K and 1500\,K (likely on the hotter end, Fig.~\ref{fig: kirchschlager2020Fig3}).  It can thus be concluded that the observed excess emission indeed stems predominantly from hot thermal emission.  Further evidence that the emission is unlikely to stem from scattered star light instead of thermal emission comes from the integral polarimetric data obtained by \citet[][see Sect.~\ref{sect_integral_polarimetry}]{marshall2016}.  Tentative evidence of a partial contribution from scattered light has been provided by the often flat spectral slopes of the excesses measured over three or seven spectral channels across the $H$~band in PIONIER data \citep{defrere2012a, ertel2014, absil2021}, however the accuracy of those measurements is limited and the slopes are still generally consistent with hot thermal emission within the error bars.  Fig.~\ref{fig_therm_sca} also illustrates that a contribution from scattered light in the $H$~and $K$~bands is plausible.

We thus conclude that the excess emission likely stems from thermal emission at a temperature of the order of 1000\,K to 2000\,K.  Thermal emission from hot dust very close to the host star is the most plausible source of this excess, though a scattered-light component at the shortest wavelength ($H$~band) cannot be ruled out and the amount of scattered light at even shorter wavelengths produced by this dust is yet largely unconstrained.

\subsubsection{Lack of mIR detections}
\label{subsec: lackOfMIRDetections}

\begin{figure}
  \centering
  \includegraphics[width=1.0\textwidth]{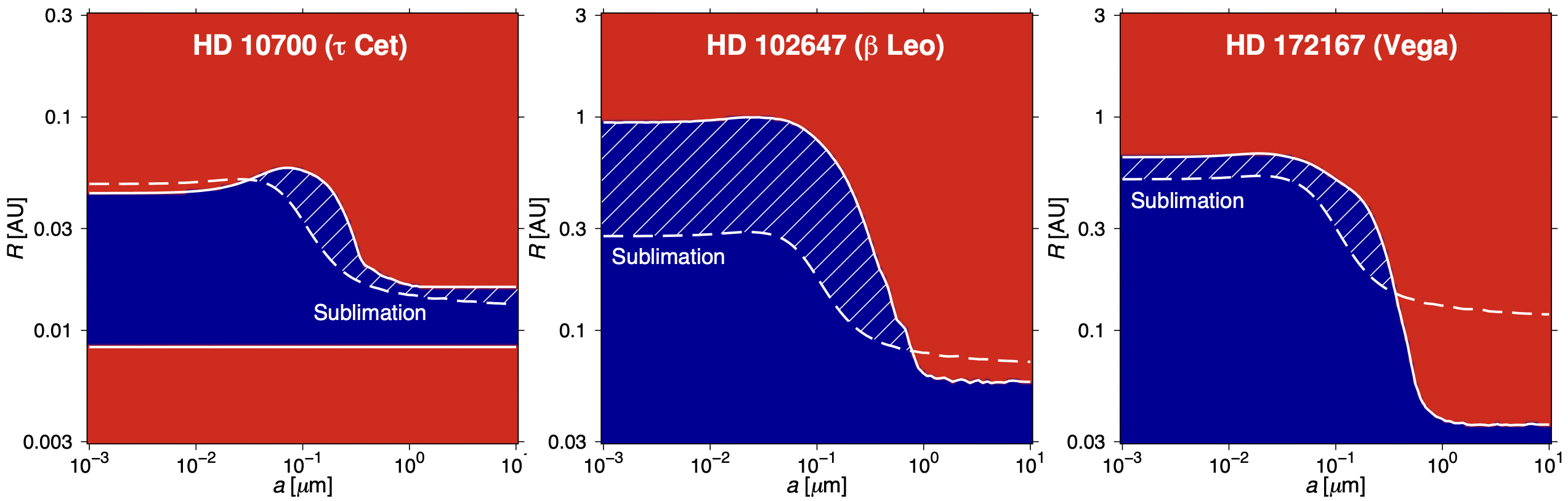}
  \caption{Fits to the nIR to mIR constraints on the shape of the emission (including mIR upper limits) for three systems from \citet{kirchschlager2017}.  The blue region shows combinations
  of grain size and separation from the host star that satisfy the observational constraints, while the dashed line shows the (grain-size dependient) dust
  sublimation radius.  Only the hashed area of typically small, hot grains is thus both physically viable and consistent with the observations.  Image reproduced from \citet{kirchschlager2017}.}
  \label{fig_kirchschlager_constraints}
\end{figure}

In the context of thermal dust emission, it is interesting to note that no hot dust has yet been conclusively detected with $N$-band nulling interferometry.  Of the nine systems \citep{kirchschlager2017} with known hot dust observed with KIN, an excess detection was achieved only around $\beta$\,Leo \citep{mennesson2014} and at limited significance around Fomalhaut \citep{mennesson2013}.  However, the KIN excess around $\beta$\,Leo is virtually identical to that detected by LBTI which is located at several au from the star \citep{defrere2021} and too cool to significantly emit in the $K$~band where hot dust has been detected with FLUOR.  The KIN excess thus likely also originates largely from this region and the contribution to the $N$-band emission from the hot dust is negligible.  For Fomalhaut, no LBTI observations are available but the KIN excess also plausibly originates from dust at a range of separations further out than the K-band-detected hot dust \citep{lebreton2013, gaspar2023}.  The upper limits on hot dust from KIN have been studied in detail by \citet{kirchschlager2017} and \citet{pearce2022}.  Their conclusions are that the dust must be both very hot and composed of very small (sub-micron) grains in order to minimize the expected $N$-band emission (Fig.~\ref{fig_kirchschlager_constraints}).  \citet{stuber2023} confirmed the need for small grains to dominate the nIR emission, but noted that a non-negligible amount of larger grains may be present in addition without violating the KIN constraints.  The models by \citet{pearce2022} further showed that there has to be a pile-up of the hottest dust grains due to some trapping mechanism to satisfy the mIR upper limits (Sect.~\ref{sec:theory}).

Of the targets observed by FLUOR \citep{absil2013}, 23 were observed by LBTI \citep{ertel2020} and of the targets observed by PIONIER \citep{ertel2014}, five were observed by LBTI.  Seven stars with near-IR detections were observed by LBTI and three of those ($\beta$\,Leo, Vega, and 110\,Her) have LBTI detections.  \citet{ertel2020} find no statistically significant correlation between the LBTI detections of warm dust and the presence of hot dust.  For the three stars with both phenomena detected, the LBTI detection is likely linked to the presence of an outer cold disk as \citet{ertel2020} find a strong correlation between the two (also see the discussion of a potential correlation between hot and cold dust below).  The four systems with detected hot dust but with non-detections from the LBTI put important constraints on the hot dust, even if they cannot resolve the near-IR emission region.  First, as discussed by \citet{pearce2022}, this means that the emission region must be inside the LBTI's inner working angle or the slope of the thermal emission spectrum must be even steeper than constrained by \citet{kirchschlager2020} due to LBTI's higher sensitivity compared to KIN.  Second, the LBTI results put even stronger constraints on the presence of dust in/near the habitable zone, migrating inwards to supply the hot dust or being blown away from the hot dust emission zone.  This may have important and strong consequences for the supply of dust to the hot exozodi (Sect.~\ref{sec:theory}).  The lack of mIR detections further shows that hot exozodiacal dust is unlikely to be composed of small silicate grains, as those would produce a major emission feature in the $N$ band and hence make a non-detection even more surprising \citep{absil2006, akeson2009, kirchschlager2017}.

\begin{figure}
  \centering
  \includegraphics[width=1.0\textwidth]{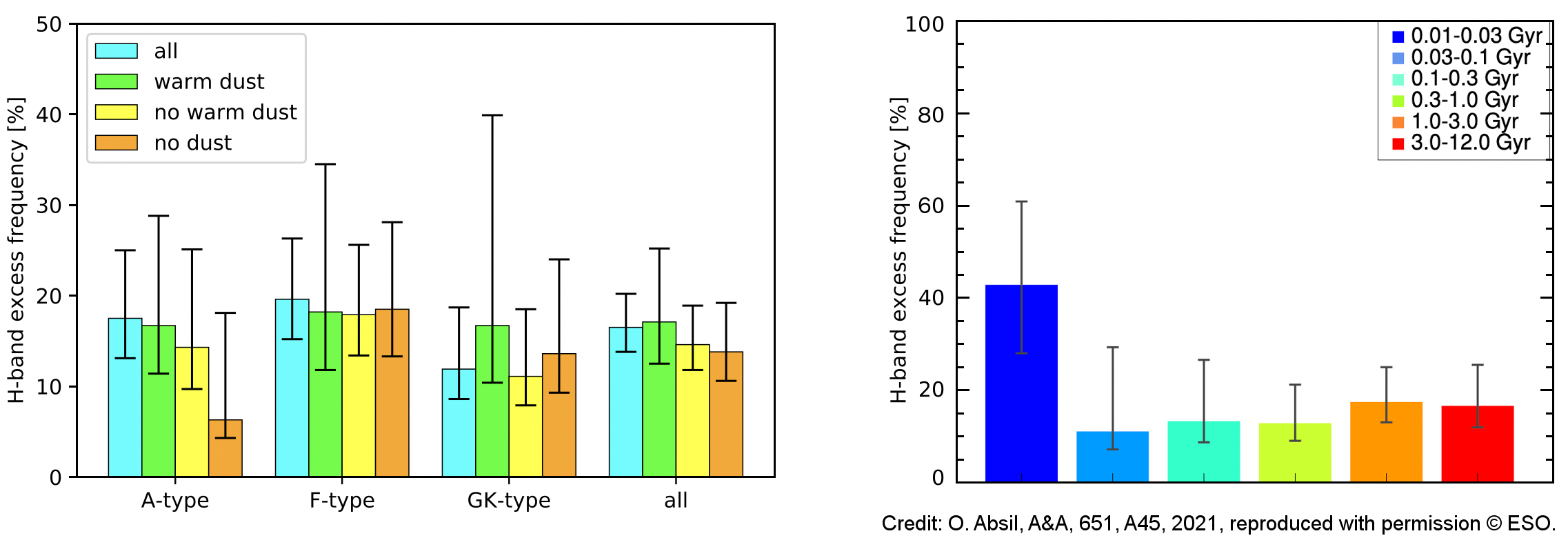}
  \caption{Detection statistics of hot dust from \citet{absil2021}.  \emph{Left:} No correlation with stellar spectral type and the presence of other
  dust in the system was found (though a potential correlation with the presence of other dust in the system was found after further analysis, see
  Sect.~\ref{sect_hot_dust_stats}).  \emph{Right:} Detection statistics with respect to stellar age.}
  \label{fig_absil2021_stats}
\end{figure}

\subsubsection{Hot dust detection statistics}
\label{sect_hot_dust_stats}
The strongest statistical constraints from exozodi-detection rates come from \citet{absil2021} due to the size of the full sample of PIONIER observations (133 stars total) analyzed.  This work shows a paucity of correlations with any other properties of the systems studied (Fig.~\ref{fig_absil2021_stats}, left).  A previously suggested correlation between the stellar spectral type and the hot-exozodi detection rate was not confirmed in the larger sample.  A correlation with the presence of an outer, cold debris disk was found for the first time, but this relies on the correction of a sensitivity bias based on assumptions on the dust location and thus still has to be considered tentative.  This lack of correlation may be related to the fact that only the brightest debris disks and hot exozodi can be detected ($\sim$20\% detection rate each).  A previously suggested correlation with the stellar age where the hot-exozodi detection rate was proposed to increase with age was not corroborated (Fig.~\ref{fig_absil2021_stats}, right), instead a higher detection rate for the very youngest systems ($<30$\,Myr) was found, which may be linked to primordial material still being present very close to the youngest stars.  Finally, no correlation was found between the detection rate of hot exozodi systems and the location of dust further out in the system where it is known (e.g., Asteroid belt vs.\ Kuiper belt analogs).  A lack of these correlations could suggest that the presence of hot dust is mostly linked to parameters of the system that are poorly accessible to us.  This could, e.g., be the specific configuration of the planetary system \citep{bonsor2012, bonsor2012a, faramaz2017, bonsor2014} if under the right configuration even small (undetectable) reservoirs of outer material could supply the hot dust, while under the wrong configuration no efficient supply is happening independent of the size of the outer reservoir.  \citet{ertel2014} found their samples to be too small to investigate potential connections between hot dust and the presence of known planets around the target stars.  This should be re-assessed given the progress on detecting planets over the past decade and the larger number of stars searched for exozodi by now.

\subsubsection{Variability}
\begin{figure}
  \centering
  \includegraphics[width=1.0\textwidth]{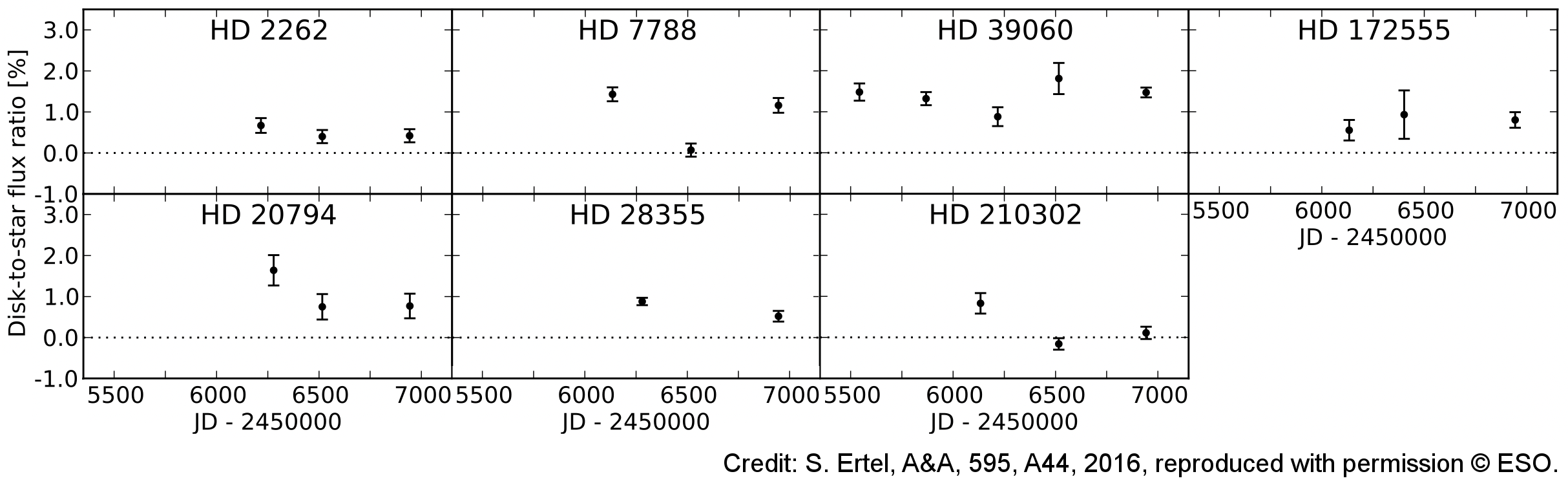}
  \caption{Time series of detected hot excesses from PIONIER \citep{ertel2016}.  One system, HD\,7788 ($\kappa$\,Tuc), shows significant variability and the excess in another system, HD\,210302, may have disappeared unless the original detection was spurious.  The other systems may also be considerably variable though no significant variability was found given the large uncertainties on the excess detections.}
  \label{fig_ertel2016_variability}
\end{figure}

\cite{ertel2016} identified at least one system, $\kappa$\,Tuc, from repeated observations that shows significantly variable excess on the time scale of one year (Fig~\ref{fig_ertel2016_variability}).  They find one additional case of potential variability and \citet{nunez2017} identified another tentative case.  This shows that the excesses can be variable, which may give important insight into the life time of the material and replenishment rates and mechanisms (e.g., frequent but major, sporadic events such as comet disruption rather than continuous inflow of dust, Sect.~\ref{sec:theory}).  Variability may also explain why not every excess is re-detected in every observation, though statistical errors close to the detection limit of the available instruments can also explain this.

\section{Theory of hot exozodiacal dust}
\label{sec:theory}

\newcommand{\timCom}[1]{\textbf{\color{red}[{\# #1 -- Tim}]}}
The generally favored explanation for nIR excesses is that they arise from very small, very hot dust grains located close to stars. Such dust would be subject to a number of physical effects, some of which would limit grain lifetimes. It is hence posited that this dust must be continually replenished, and/or be subject to some mechanism that allows it to survive and remain near the star. However, dynamical explanations of hot dust are challenging, and to date no model has self-consistently explained hot exozodi for the full range of stars about which it is observed. In this section, we discuss the main challenges for hot-exozodi models, and summarize the major models so far. We end this section by discussing how hot-exozodi knowledge may be leveraged to retrieve information on the architecture of the associated planetary system.

\subsection{The exozodiacal-dust survival challenge}

The fact that hot exozodi is detected at all is difficult to explain. Any dynamical model must contend with two physical processes: blowout and sublimation. Each of these processes has the potential to severely limit hot-dust lifetimes, removing grains on timescales that can be as short as days or even hours. Any hot-exozodi model must account for both of these processes, which pose significant challenges to overcome. Here we explain these two processes in more detail.

Like dust in debris disks, exozodiacal dust is subject to radiation forces. Small grains are influenced by radiation pressure from photons, which manifests as an outward force that counteracts the inward pull of gravity.  The relative strength of radiation pressure versus gravity is characterized by the parameter $\beta$, which depends on grain size, material and star type. Since both forces scale as distance$^{-2}$, radiation pressure effectively reduces the gravitational force from the star by a factor ${(1-\beta)}$ (e.g. \citealt{gustafson1994}).  The gravitational force scales with the third power of the grain size (with grain mass, i.e., volume), while the radiation pressure in the geometric regime (for grains larger than a few micron) scales with the second power or the grain size (with cross section).  Hence, radiation pressure is thought to be significant for the small, sub-micron sized grains inferred from nIR observations.  This leads to the concept of the \mbox{\textit{blowout size}}; for grains smaller than this size, radiation pressure is significant enough to overcome the force of gravity, and the grain would quickly be blown out of the system. For example, a ${0.1\; \mu}$m carbon grain released by a star-grazing comet at 0.25\,au from an A0V star would take just 0.01\,years to blow out to 10\,au \citep{pearce2022}. This leads to a significant problem; many hot exozodi systems are thought to be comprised of sub-micron grains that are smaller than the blowout size (depending on material), so it is very difficult to explain how such grains are present in large quantities near stars. The blowout size varies with stellar type, with more-luminous stars having larger blowout sizes, so this is a significant problem for A-type stars in particular.

In addition to radiation pressure, hot dust close to stars is also subject to sublimation. The temperature at which sublimation becomes significant depends on the dust composition, but is roughly \mbox{2000 K} for carbon and \mbox{1200 K} for silicates \citep{kobayashi2009}. Grains near the sublimation temperature would rapidly turn into gas. The inferred temperatures of observed hot exozodi are often within or near these sublimation regimes (e.g. \citealt{kirchschlager2017}), so sublimation may be a significant factor in hot-dust evolution. Like radiation pressure, sublimation could severely limit the lifetime of hot dust.

Hot-dust survival is further compounded by its proximity to stars, which would result in very short orbital periods for an \textit{in-situ} population. These periods would be of the order of days for typical hot-exozodi parameters, meaning that dynamical timescales would also be very short. One consequence of this is that, in addition to radiation pressure and sublimation, hot dust may also have to contend with rapid collisional erosion. A second consequence is that hot dust cannot be sustained via a collisional cascade in an \textit{in-situ} planetesimal belt, unlike `cold' debris disks located farther from stars, because such a belt would quickly collisionally erode \citep{wyatt2007}. For example, a population of km-sized planetesimals massive enough to produce inferred hot-dust levels (up to $\sim$$10^{-6}\,\mathrm{M}_\oplus$; \citealt{kirchschlager2017}), and located at $\lesssim$1\,au, should only survive for several hundred years before collisionally depleting \citep{kral2017}. The probability of observing such a system at the right time would be extremely low, and incompatible with the observed detection rate of $\sim$20\%. 

Given the above challenges, if we are to explain nIR excesses as hot dust then we need some other mechanism to replenish and/or sustain dust. This mechanism is currently unknown. Many dynamical models have been suggested, and these can be roughly divided into two groups: `supply only' models, where dust lost through the above processes is continually replenished by some means, and `trapping' models, where some additional mechanism protects grains from blowout and/or sublimation. The following sections discuss the major models in each category. 

\subsection{Supply only models \& their limitations}

There are two main dust-supply mechanisms that are commonly discussed in hot-exozodi literature. These are supply via Poynting-Robertson (PR) drag from a distant debris belt, and supply via comets. A third possibility, disintegrating close-in planets, has also been considered. From current models, it appears that no supply mechanism can, by itself, reproduce all constraints on hot exozodi. In this section we summarize these mechanisms and their limitations. 

\subsubsection{Poynting-Robertson drag}

\begin{figure}
  \centering
  \includegraphics[width=8cm]{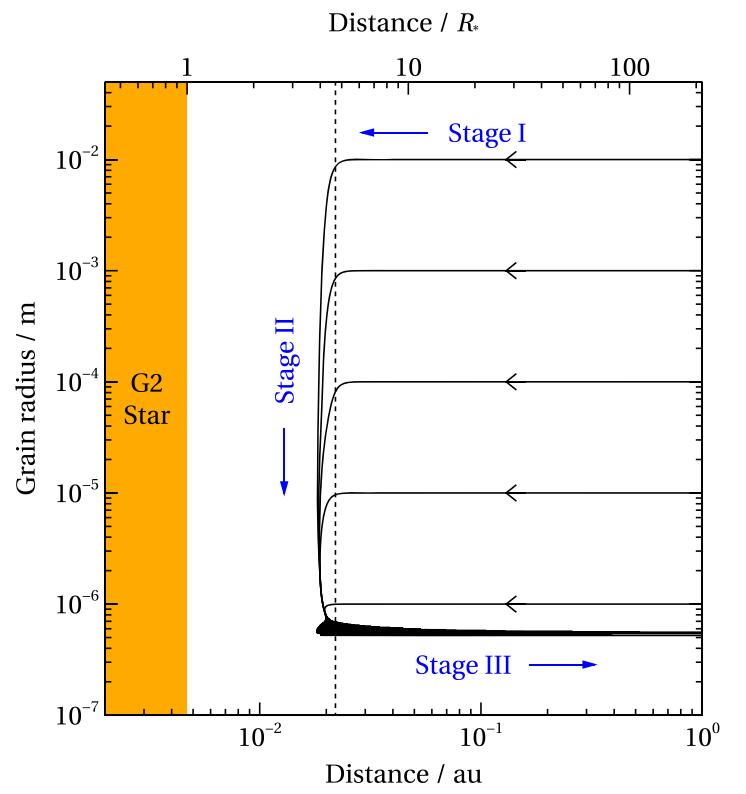}
  \caption{Evolution of dust grains in the collisionless PR-drag supply model by \citet{pearce2020}. Black lines show the evolution of five carbon grains with different initial sizes. Grains first migrate inwards under PR drag (Stage I), before reaching the sublimation region (dashed line) and rapidly sublimating (Stage II). Grains shrink until they approach the blowout size, where their orbits become increasingly eccentric and eventually they get blown out of the system (Stage III).  Image reproduced from \citet{pearce2020}.}
  \label{fig: pearce2020Fig1}
\end{figure}

PR drag is a relativistic effect and a second manifestation of the same force that causes radiation pressure, arising from interactions between dust grains and stellar photons.  It causes dust to gradually lose angular momentum and spiral inwards onto the star \citep{poynting1904,robertson1937}. The PR-drag model for hot exozodi postulates that dust originates in some distant debris disk, such as an asteroid belt in the terrestrial region, and then spirals inwards towards the star under PR drag. As the dust gets closer to the star, it heats up. If the grain does not get destroyed by collisions or ejected by planets, then eventually it gets close enough (and hot enough) that it approaches its sublimation temperature. At this point a dust grain would rapidly shrink through sublimation, and as it shrinks, the relative effect of radiation pressure becomes more and more significant. This causes the orbit of the grain to become more and more eccentric. Eventually, the shrinking grain reaches the blowout size; at this point, radiation pressure is sufficient to increase its eccentricity to 1, and the grain is blown out of the system. The general PR-drag model has been explored multiple times in the literature (e.g. \citealt{krivov1998, kobayashi2008, kobayashi2009, vanlieshout2014, sezestre2019, pearce2020}), and is shown on Fig. \ref{fig: pearce2020Fig1}.

Unfortunately the PR-drag model alone cannot reproduce the observational constraints on hot exozodi, for two reasons. First, the model produces too much mIR emission; observed hot exozodi appears to have significant nIR emission yet no detected mIR emission (Section \ref{subsec: lackOfMIRDetections}), and this constraint is violated by the PR-drag model. The violation arises because dust in the PR-drag model would spend the vast majority of its time away from the star, where it is cooler and emits in the mIR, and only a tiny fraction of its time as very small, very hot grains emitting in the nIR before getting blown away. This is the result of the PR-drag timescale being very long, much longer than a grain's orbital period, whilst the sublimation and blowout timescales are much shorter. Hence dust in the PR-drag model emits far too much in the mIR compared to the nIR.

The second reason why the PR-drag model fails is that it cannot produce dust smaller than the blowout size. This is not a problem for Solar-type stars, but it is for A-type stars, which have larger blowout sizes (${5\;\mu}$m for solid carbon grains around an A0V star, versus ${0.5\;\mu}$m for a G2V star; \citealt{pearce2020}). This means that, for A-type stars, grains in the PR-drag model simply cannot get hot enough or small enough to produce sufficient nIR emission before being blown away. Taken together, these two reasons mean that the PR-drag supply model alone cannot reproduce hot-exozodi observations, and the problem is even more acute for A-type stars than for Solar-type stars. PR supply is also further hampered if dust collisions are considered; in this case, much of the inflowing material may be destroyed by collisions before reaching the sublimation region \citep{vanlieshout2014}.

\subsubsection{Cometary supply}

\begin{figure}
  \centering
  \includegraphics[width=12cm]{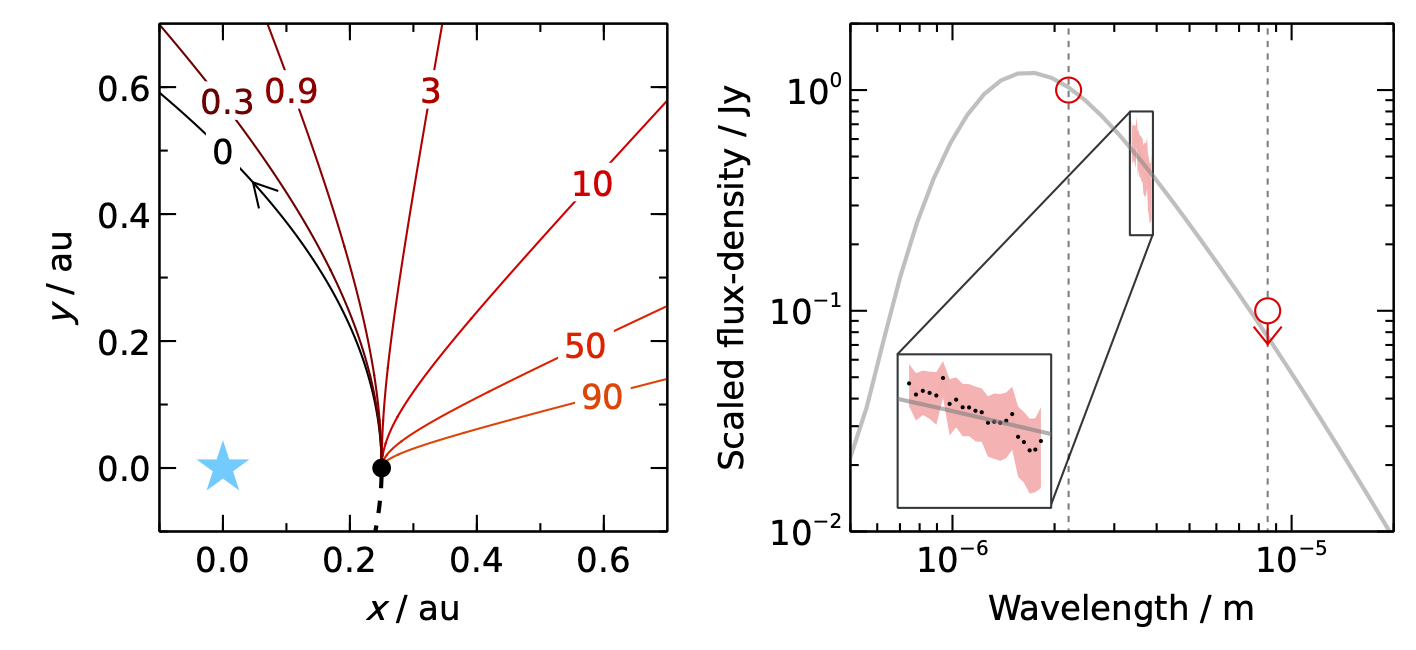}
  \caption{\emph{Left:} Dust is released by a star-grazing comet at pericenter (black circle). Different grain sizes have different trajectories, owing to the different effects of radiation pressure (colored lines). The comet has pericenter 0.25\,au and eccentricity 0.999, and travels in the direction indicated by the arrow. The star's spectral type is A0\,V. The $\beta$ value of each grain is shown on its trajectory. \emph{Right:} Generic hot exozodi with typical detected flux at 2.2\,$\mu$m and a typical upper limit at 8.5\,$\mu$m (red circles).  The VLTI/MASTISSE data of $\kappa$\,Tuc from \citet{kirchschlager2020} are scaled and overplotted as well for comparison.  Under certain assumptions, the resulting model (gray line) can reproduce hot exozodi observations. However, these assumptions are difficult to physically justify.  Figure adapted from \citet{pearce2022}.}
 \label{fig: pearce2022Fig2}
\end{figure}

Since the PR-drag model fails, partly because grains produce too much mIR emission as they slowly spiral towards the star, a more promising alternative is cometary supply. In this model, comets make close approaches to a star, and deliver dust at each pericenter passage (Fig.~\ref{fig: pearce2022Fig2}, left). An advantage of this model over PR supply is that dust can be delivered directly to the hot-emission region, which significantly reduces mIR emission by minimizing the time that individual grains spend away from the star \citep{sezestre2019}.

There are several other advantages of cometary-supply scenario. First, stochastic cometary inflow may be the simplest way to explain nIR variability, which is seen in at least one system \citep{ertel2016}. Star-grazing comets with the required orbits are present in the Solar System (e.g. \citealt{Kreutz1888, Opik1966, Marsden1967, Marsden1989, Marsden2005}), and there is evidence of exocometary activity in other systems; this is in the form of suddenly appearing and slowly disappearing metallic lines \citep{Ferlet1987, lagrange1988, beust1989}, variable spectral features \citep{Kiefer2014, rebollido2020, rebollido2023} and transits \citep{LecavelierDesEtangs2022}. Much of the detected exocometary activity is around the star \mbox{$\beta$ Pic}, which notably also hosts a nIR excess \citep{defrere2012b}. In addition, there may be a tentative relation between nIR excesses and circumstellar gas indicative of cometary activity \citep{rebollido2020}. For these reasons, cometary supply may appear to be a more promising nIR-generation mechanism than PR-drag supply.

However, despite these advantages, pure cometary supply seems unlikely to be the mechanism generating hot exozodi. This is because the deposited dust should quickly be removed from the hot-emission region close to the star, either by sublimating, colliding or blowing away under radiation pressure. If the deposited dust blows away, then it would produce too much mIR emission as it cools, which is similar to the main problem in the PR-drag model \citep{sezestre2019}. The only known setup that reproduces hot-exozodi observations via cometary supply for both Sun-like and A-type stars is for comets to deposit dust so close to stars that the grains sublimate before they can blow away \citep{pearce2022}. However, this mechanism appears unlikely, for two main reasons. First, it requires unreasonably huge dust-deposition rates to counteract the extremely short grain lifetimes (Pearce2022 found that dust-input rates of $10^{-6}$ to $10^{-4}$ M$_\oplus$/yr were required, equivalent to the total disintegration of 20 to 2000 comets of radius 50 km per year). Second, it requires the initial size distribution of grains ejected by exocomets to be much steeper than those of Solar-System comets. A third issue is cometary origin; whilst many mechanisms can produce star-grazing comets from a distant debris disk (e.g. \citealt{bailey1992, bonsor2012, bonsor2013, bonsor2014, faramaz2017, marino2018}), there is no significant correlation between nIR excesses and those in the mIR or fIR that would indicate a cometary source \citep{millan-gabet2011, ertel2014, ertel2018, ertel2020, mennesson2014, nunez2017, absil2021}. For these reasons, \cite{pearce2022} conclude that cometary supply alone is unlikely to be the mechanism that generates hot exozodi.

\subsubsection{Disintegrating inner planets}

Another possibility is that close-in, disintegrating planets are the hot-dust source \citep[e.g.,][]{rappaport2012}. This is one scenario that \citet{lebreton2013} considered to explain Fomalhaut's nIR excess, although they disfavored it in that case. The authors felt it was unclear whether a disintegrating or evaporating planet could release enough dust to sustain a hot exozodi, and also that the dust distribution produced by a planet would be very different to that inferred for the nIR excess. More generally, since hot exozodi are detected around ${20\%}$ of main-sequence stars, with very diverse ages and spectral types, it is potentially difficult to envisage that all of these systems host a disintegrating inner planet at the time of observation. Furthermore, no correlations between nIR excesses and detected planets were found by \cite{ertel2014}, although as exoplanet-detection techniques have since increased in sensitivity this should be reassessed. It is also worth noting that the disintegrating-planet scenario has been much less explored than the PR-drag and cometary supply models, and so additional work here may be warranted. Like the cometary supply model, dust released from a disintegrating planet would probably have to sublimate before blowing away if it is to reproduce hot-exozodi observations, to prevent emission being dominated by cooler grains on their way out of the system.

\subsubsection{General limitations of supply only models}

The fundamental problems faced by the PR-drag and cometary models are common to all other supply models. Regardless of how dust of the required size gets deposited close to a star, those grains should only survive for very short timescales before sublimating or blowing away. The result is that huge dust-inflow rates would be required to sustain the observed nIR excesses in these models, which leads to problems with both physical plausibility and the lack of mIR and fIR excesses in many systems. Generally, current modelling suggests that simply getting dust close to a star is not enough to reproduce hot-exozodi observations. This has led to the hypothesis that some additional mechanism operates in the inner regions of these systems, which we discuss in the following section.

\subsection{Trapping models \& their limitations}

Simply getting dust to the innermost regions of planetary systems appears insufficient to generate hot exozodi, because these very hot, very small grains should not survive for long. This problem has motivated a second class of hot-exozodi models, collectively known as `trapping' models. Their fundamental idea is that some mechanism greatly extends the amount of time that dust can spend near stars, which increases the overall nIR-to-mIR-emission ratio and reduces the required dust-inflow rate. However, current trapping models also have problems, and to date none has successfully reproduced hot-exozodi observations across all star types. In this section we summarize the main trapping models and their limitations.

\subsubsection{Magnetic trapping}

\begin{figure}
  \centering
  \includegraphics[width=8cm]{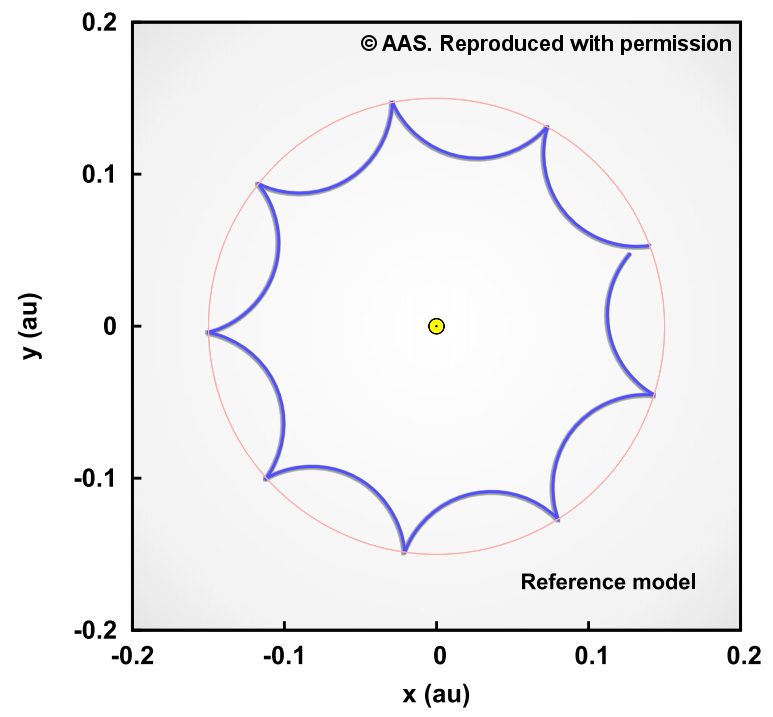}
  \caption{Magnetic-trapping model \citep{rieke2016}. A charged dust grain gets trapped in stellar magnetic fields, and follows a trajectory shown by the blue line.  In this model, the grain is released at the sublimation radius (red line), and the magnetic field rotates with the star, in the same direction as the grain's orbit.  Image reproduced from \citet{rieke2016}.}
  \label{fig: rieke2016Fig2}
\end{figure}

The concept of magnetic trapping is that charged dust grains get trapped in stellar magnetic fields. The idea is that this trapping resists the outward force of radiation pressure, and therefore allows dust grains smaller than the blowout size to remain close to the star (Fig. \ref{fig: rieke2016Fig2}; \citealt{czechowski2010, su2013, rieke2016, stamm2019, kimura2020}).

Dust grains in the stellar environment can quickly become charged through various mechanisms, including the photoelectric effect and the plasma current (e.g. \citealt{belton1966, lefevre1975, draine1979, kimura1998, rieke2016, kimura2020}). Provided the charged grains are small enough (${\lesssim 100 \; \rm nm}$ for an A-type star; \citealt{rieke2016, kimura2020}), and located close enough to the star, they could potentially get trapped in stellar magnetic fields. Such grains released from initially circular orbits could undergo long-term epicyclic motion around the star, even under the influence of radiation pressure. In theory, this mechanism could significantly extend the time that small grains spend in the hot-emission region.

However, there are problems with current magnetic-trapping models. One issue is that, whilst trapped grains would not blow out of the system, there is nothing protecting them from sublimation. Hence current magnetic models may not significantly extend grain lifetimes very close to the star, where sublimation would be the dominant process \citep{kimura2020}.  Whilst it may be possible to trap grains outside the sublimation region, for example if the magnetic field rotates in the opposite direction to the grain's orbit \citep{rieke2016}, nIR excesses imply the grains should be very close to the sublimation region and hence susceptible to sublimation. Another issue is that, for A-type stars at least, magnetic trapping is only efficient for grains smaller than the blowout size. Hence sub-blowout grains must somehow get into the magnetic-field region without first blowing away \citep{kral2017}, and no model has yet demonstrated a means to do this (although star-grazing comets are a possibility). A third problem, and perhaps the most important, is that nIR excesses are detected around stars with very diverse spectral types, magnetic-field strengths and rotation speeds, all of which should affect the efficiency of magnetic trapping \citep{kimura2020}. The lack of any clear correlations between these stellar properties and the presence of nIR excesses may disfavour magnetic trapping as the hot-exozodi mechanism. However, further investigations into magnetic trapping are currently ongoing (Peronne et al., in prep.).

\subsubsection{Gas trapping}

\begin{figure}
  \centering
  \includegraphics[width=8cm]{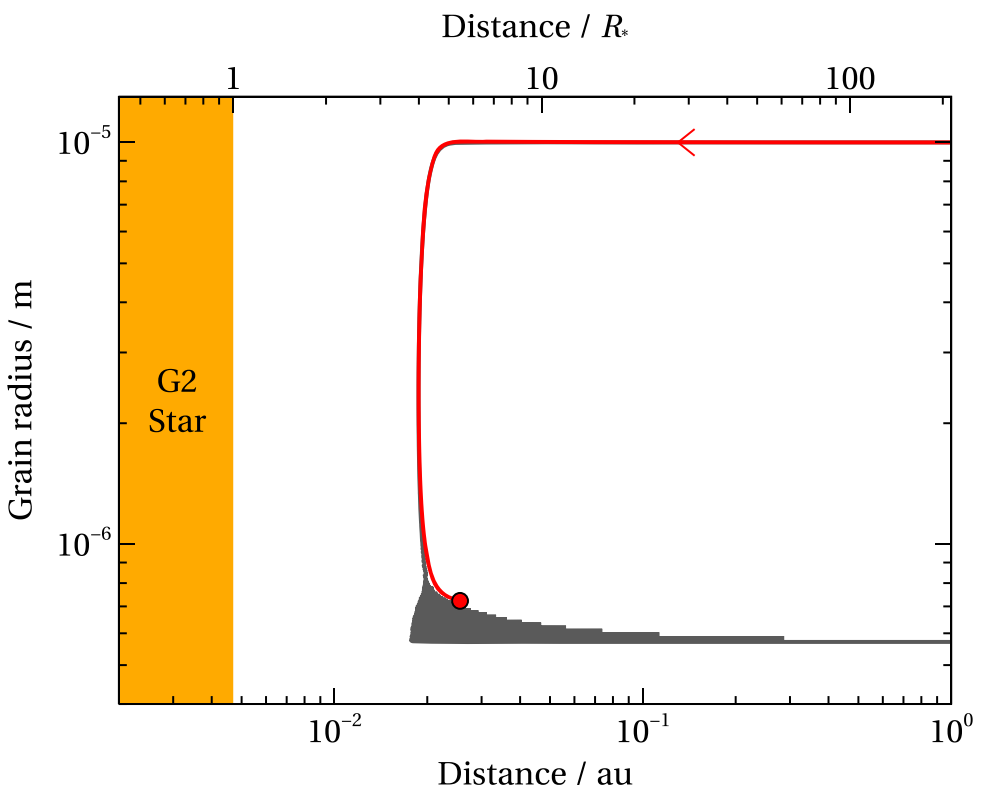}
  \caption{Evolution of a dust grain in the gas-trap model (red line) by \citet{pearce2020}. The grain migrates inwards through PR-drag and sublimates, like in the PR-drag model on Fig. \ref{fig: pearce2020Fig1}. However, as the grain shrinks it becomes coupled to the gas released by sublimation. This traps the grain immediately outside the sublimation region. The gray line is the gas-free case for comparison.  Image reproduced from \citet{pearce2020}.}
  \label{fig: pearce2020Fig3}
\end{figure}

An alternative mechanism is gas trapping. In this model, dust that is supplied by either PR drag or comets comes close to the star and sublimates, releasing gas. This gas then traps subsequent incoming grains, and it is these hot grains that produce the nIR emission (Fig. \ref{fig: pearce2020Fig3}; \citealt{lebreton2013, pearce2020}).

The gas-trapping model has several advantages. It naturally traps grains just outside the sublimation region, where nIR excesses are thought to originate; this is consistent with the inferred scaling of hot-dust location with stellar luminosity \citep{kirchschlager2017}. It is also independent of stellar parameters such as spectral type, age and rotation rate, so could potentially operate for any star with sufficient dust inflow. The model also predicts trapped grains to have a steep size distribution, like those inferred for hot dust, because small grains would be trapped more efficiently. Finally, the mechanism may be quite efficient, with models requiring only small quantities of gas that are difficult to rule out observationally \citep{pearce2020}.

However, like all other attempts to explain hot exozodi, the gas-trapping model does not fully explain all observable features of hot exozodiacal dust. The most significant issue is that, unlike magnetic trapping, it is very difficult to trap sub-blowout-sized grains in gas \citep{pearce2020}. This is not a problem for Sun-like stars, where the observed nIR excesses can arise from grains larger than the blowout size, but it poses a major problem for A-type stars which require sub-blowout grains \citep{kirchschlager2017}. Current models are also relatively simple, assuming that gas released by grain sublimation spreads into a continuous radial profile similar to a scaled-down protoplanetary disc, but complicating effects like sporadic gas injection, photoevaporation, vertical spreading and turbulence have not been explored.

\subsubsection{Differential Doppler Effect (DDE)}

Another potential trapping mechanism is the Differential Doppler Effect. This is a drag force arising from the Doppler shifting of incident radiation on dust grains \citep{burns1979}. However, it appears that this mechanism cannot generate sufficient drag to trap small grains; \cite{sezestre2019} showed that\, for realistic star-rotation rates, this effect is only significant compared to PR drag for grains with ${\beta \sim 1}$.  Such grains should rapidly blow away under radiation pressure, which is much stronger than DDE, so DDE appears to be ineffective at trapping hot dust.

\subsubsection{General limitations of trapping models}

There are two main issues that pose difficulties for any trapping model. The first is that hot-dust lifetimes should be affected by both blowout and sublimation; a trapping model cannot completely solve the hot-exozodi problem unless it addresses \textit{both} of these processes. For example, the current magnetic-trapping models could potentially trap sub-blowout grains, but it is unclear whether those grains would simply sublimate instead. Likewise, whilst the gas-trap model naturally protects grains from sublimation, it is unclear whether it can trap grains smaller than the blowout size. These two competing requirements make it very difficult to devise a model that can effectively trap and retain very hot, sub-blowout-sized dust, yet these are the grains inferred from nIR excesses around A-type stars (it is easier for Sun-like stars, where the blowout size is smaller due to reduced radiation pressure and hence the nIR excesses can be reproduced without requiring sub-blowout grains).

The second issue is nIR variability. The fundamental motivation for trapping models is to significantly increase the amount of time that hot dust can survive close to a star, with the aim of increasing the nIR-to-mIR-emission ratio and reducing the required dust-inflow rate. Hence the longer a grain could be trapped, the more efficient and advantageous a trapping model becomes. However, at least one nIR excess is seen to repeatedly vary by 100\% (a few 10\% minimum within the error bars), over at most a $\sim$1\,year timescale \citep{ertel2016}. Such high variability can be difficult to reconcile with long-term trapping models, since the quantity or properties of trapped dust would have to vary significantly. Rapid variability would also imply that, if trapping mechanisms operate in nature, their effectiveness is more limited than some theoretical models require.

\subsection{Possible solutions and future modeling efforts}

No theoretical model has self-consistently explained all hot-exozodi observations. Some models, such as PR-drag supply, do not appear to work in any case \citep{sezestre2019}; other models, such as gas trapping, can reproduce observations for some star types but not others \citep{pearce2020}. The difficulty theorists have had in reproducing these observations suggests that we are either omitting some key physical processes, or that current modeling is built on at least one incorrect assumption. Below we list several possibilities that we believe should be considered in future modeling efforts; investigations into some of these are planned in upcoming work.

\subsubsection{Is there a meaningful contribution from scattered light to the nIR excesses?}

As shown in Fig.~\ref{fig_therm_sca}, a contribution from scattered light to the excess emission in the nIR interferometric data is plausible.  This could distort the color and hence the temperature measurements, making the dust appear hotter than it is.  If the dust was cooler, and hence further away from the host star, this could at least partially mitigate some of the challenges in explaining the hot-dust phenomenon.  However, it is important to note that other constraints such as the lack of mIR emission from the dust observed by KIN or LBTI strongly limits the plausible range of separations at which the dust could be located.  Typically less stringent but more direct constraints on the location of the nIR-detected dust are also imposed by the small field of view of the nIR-interferometric instruments (a few 100\,mas in radius).  \citet{stark2015} discuss the case of edge-on systems with cold debris disks, but is appears implausible that all detected hot-dust systems are caused by extreme forward-scattering in outer, cold debris disks seen edge-on.  Integral polarization measurements \citep{marshall2016} do not strongly inform the amount of scattered light from the hot dust due to degeneracies between dust geometry and the degree at which individual grains polarize the scattered light (Sect.~\ref{sect_integral_polarimetry}), though they generally suggest low scattered-light content at visible wavelengths.  Nonetheless, detecting the dust at shorter nIR and visible wavelengths, i.e., shortward of the $H$~band, or even placing meaningful upper limits on its brightness, would be useful and enhance the constraints from already available nIR data.

\subsubsection{Are the assumed grain-physics prescriptions accurate?}

Any hot-dust model must include prescriptions for grain dynamics, sublimation and emission. Typical models assume: Newtonian gravity, standard radiation-pressure and PR-drag prescriptions, sometimes additional forces like magnetic-field interactions, prescriptions for continuous sublimation, and often Mie theory to describe dust temperature and emission (e.g. \citealt{lebreton2013, vanlieshout2014, kimura2020, sezestre2019, pearce2020, pearce2022}). However, these models are then applied to an extreme scenario: very small, very hot grains located very close to stars. It is therefore pertinent to ask whether these prescriptions accurately model the required physics.

For example, some models assume nanometer-sized grains, which would be just ${\sim 10}$ atoms wide. This dust would be immersed in an intense radiation field, and so may be ionized. Would the above sublimation and emission prescriptions be valid for such grains?  Would stochastic heating by individual photons be a dominant emission mechanism?  Would additional processes like sputtering affect the results? It is possible that we are struggling to reproduce hot-exozodi observations because current models do not sufficiently capture the required physics.

\subsubsection{Are we considering the right dust composition?}

Most hot-dust models consider grains to be some form of carbonaceous material. This is because carbon is the only tested material that can both survive at the inferred distances and produce at least 10 times more nIR emission than mIR, as required by observations \citep{absil2006, diFolco2007, akeson2009, lebreton2013, kirchschlager2017, sezestre2019}. In particular, silicates are disfavored due to their strong mIR-emission features. However, only a few potential hot-dust materials have actually been modeled, so it is possible that some other material could reproduce observations whilst, for example, better resisting sublimation and/or blowout. The specific material need not be common throughout the whole system; if dust supply and removal rates were sufficiently high, and one specific material could survive much longer than others, then even a rare material could build up to significant amounts near the star whilst all others were quickly removed. This leaves open the possibility that some exotic, untested material could be responsible for the nIR emission.  Such grains could also produce yet-unexplored emission features that could make the excesses appear brighter and steeper over the probed wavelength regime than pure blackbody emission.  In addition, modeling complex grain compositions and shapes is computationally expensive and thus reliant on simplifications.  This includes the dynamics of highly porous/fractal grains under radiation pressure and the outcomes of their collisions and sublimation.  Our understanding of such effects is incomplete and requires more research.

\subsubsection{Are data taken at different times compatible?}

Much of the observational data on hot exozodi are fluxes in individual wavebands from several instruments, each at different wavelengths and potentially taken years apart. Usually, theorists try to fit all these data simultaneously with a dynamical model. However, since some excesses appear to be highly variable, it is not clear that this is a valid approach. For example, \citet{kirchschlager2020} took mIR data of \mbox{$\kappa$ Tuc} with MATISSE ($\sim3.5\,\mu$m), but this exozodi is known to vary by 100\% (a few 10\% minimum within the error bars) in the nIR ($1.6\,\mu$m, \citealt{ertel2016}). Therefore, depending on which nIR flux state is assumed at the time of the mIR observation, the instantaneous hot-exozodi SED could have many different shapes (Fig. \ref{fig: kirchschlager2020Fig3}). It is therefore possible that the emission from hot exozodi may vary simultaneously in both nIR and mIR, but that nIR and mIR observations, taken at different times, correspond to incompatible flux states.

There are several ways to counter this possibility. First, spectrally-resolved data from a single observation would constrain the instantaneous spectral slope across multiple wavelengths.  MATISSE is just starting to do this in the $L$~band; for $\kappa$\,Tuc, the hot-exozodi excess has a clear spectral slope in this waveband \citep{kirchschlager2020}. This slope is consistent with those inferred for other systems from combinations of nIR and mIR observations in individual wavebands (Fig. 2 in \citealt{pearce2022}), suggesting that these inferred slopes are real.  Second, data could be taken at similar times on several instruments in different individual wave bands. This would require coordinated instrument programs, but if accomplished, it would provide reliably compatible multi-wavelength data that could be safely modeled.

On the other hand, it is important to note that unequivocal variability of the nIR excess has so far only been seen for one system \citep{ertel2016}.  Given that so much of our understanding of hot exozodiacal dust hinges on the question whether or not the excess is significantly variable, this result needs to be confirmed.  Alternative explanations for the varying signal such as spatial variation from a moving clump or the likely rare possibility of a binary companion mis-identified as exozodi need to be explored.  If confirmed, it needs to be understood how common variability really is, and what the typical variability amplitudes and time scales are.

\subsubsection{Is just one mechanism responsible?}

The tendency in the literature is to assume that all hot exozodi systems are generated by a single universal mechanism. However, this makes them very difficult to explain, because that one mechanism would have to produce hot exozodi in a wide range of systems with very different stellar types, ages, and warm- and cold-dust levels. An alternative possibility is that different systems have different hot-exozodi production mechanisms; for example, it could be that gas trapping operates around some stars, whilst magnetic trapping operates around others. Whilst this multi-mechanism solution is less elegant than a single, universal process, and could be liable to overfitting if a large number of potential mechanisms were considered, it is nonetheless possible. We could potentially test this scenario in the future by acquiring detailed hot-exozodi data for a large number of diverse systems.  For example, if we could show that hot exozodi has fundamentally different properties around, say, young A-type stars compared to old G-type stars, that cannot be easily linked to the difference in stellar spectral type alone such as blowout size, then this could favor a multi-mechanism solution.

\subsubsection{Is some other, unconsidered dust mechanism responsible?}

Since only a few hot-exozodi mechanisms have been proposed and tested, it is quite possible that some unconsidered means of supplying and/or trapping dust is actually responsible. New dynamical theories are often difficult to construct; since the observational data is very sparse (usually detections in just one or a few wavebands per system), new models tend to be conceptualized first, then tested to see whether they are consistent with the little data we have. However, as new instruments come online and hot-exozodi data become more detailed, theorists will have more information on dust location, morphology and composition. This will hopefully present a clearer picture of what is happening in these systems, which could then motivate and drive new dynamical theories.

\subsubsection{Are we \textit{sure} that nIR excesses stem from dust?}

Interferometric detections of ${\sim 1\%}$ visibility deficits in the nIR suggest that something bright and extended exists in the innermost regions of planetary systems. The assumption is that these are populations of hot dust, but since we have so far failed to reproduce observations with hot-dust models, it is pertinent to ask whether this assumption is correct.

Several alternative, non-dust scenarios have been considered in the literature, and these have generally been ruled out. For example, the excesses are unlikely to be all caused by very close companions to the target stars.  The companions would generally have to be of stellar nature to explain the 1\% excess.  These are expected to produce detectable radial velocity signals, except for orbital viewing angles very close to face-on, and to generally produce astrometric signals.  While close to the sensitivity limit for the required companions, a lack of interferometric closure-phase signal also suggests that companions are an unlikely source of the detections \citep{ertel2014, marion2014}.  Although a tentative closure-phase signal was recently found (Stuber et al., in prep.) in new observations of a system with known excess, it seems unlikely that 10-20\% of main-sequence stars host very close, stellar companions and this hasn't been noticed before.  A more plausible explanation for the closure phase signal may be significant structure in the detected dust distribution.  Interestingly, nulling interferometry at both Keck and LBTI would be sufficiently sensitive to detect the companions responsible for the $H$- and $K$-band signals in the separation ranges they cover, but those data don't usually show any counterpart to the nIR detections \citep{mennesson2014, ertel2018, ertel2020}.  Similarly, whilst a non-spherical star could produce the visibility deficit, the stellar oblateness would have to be significant and would often contradict other measurements \citep{absil2006}.

However, it is still possible that these signals arise from other, unconsidered sources. For example, would extended, transient phenomena like coronal-mass ejections reproduce nIR excesses?  What would the Solar corona look like through a distant interferometer at different phases of the Solar cycle?  Could extreme forward-scattering of interstellar-medium dust along the line of sight explain the phenomenon?  Free-free emission from stellar mass loss has been partially explored and unpublished results suggest that unreasonable mass-loss rates that can otherwise be ruled out observationally would be required, but this scenario should be explored more formally and extensively.  As hot dust appears so difficult to reproduce theoretically, it may be time to start asking whether we really have exhausted all alternative scenarios.


\subsection{Connection to planetary systems}

As described in the previous sections, the origin of the hot dust is so far poorly understood.  What seems clear is that the material cannot exist in the location where we observe it for the life time of the stars around which it is detected.  Hence, any explanation of the presence of the dust requires at the very least two basic ingredients, a source of the material and a mechanism to deliver the material to the innermost regions of the system.  As an interstellar source seems implausible, the source of the hot dust must almost certainly be circumstellar, such as a circumstellar dust disk, a planetesimal belt such as the Solar system's Asteroid belt or Edgeworth-Kuiper belt, or a cloud of minor bodies such as the Oort cloud.  The delivery of the material to the hot-dust region then plausibly involves its interaction with planets in the system, be it as a mechanism for the delivery or as an obstacle to it.  Hence, it seems likely that in one way or the other the presence, properties, and location of the hot dust are tied to the architecture, dynamics, and evolution history of the planetary system that harbors it.

It is this line of arguments that has motivated dynamical studies of potential delivery mechanisms for hot (and warm) exozodiacal dust based on planet-disk interaction.  The delivery of material depending on the architecture of a planetary system and planetesimal belt has been studied, including the effects of the detailed mass and orbital configuration of a multi-planet system \citep{bonsor2012, bonsor2012a}, the addition of a major dynamical instability of a planetary system \citep{bonsor2013}, and the potential of planetesimal-driven migration of the outer-most planet in the chain \citep{bonsor2014}.  Perhaps the most promising scenario was presented by \citet{faramaz2017} with a configuration of a single giant planet orbiting outside a planetesimal belt.  It is conceivable that such a scenario similar to our Solar system's configuration with Jupiter and the Asteroid belt is relatively common and this has been shown to be able to produce comets that reach sufficiently close to the host star over Gyr time scales.  All these scenarios are found to be sensitive to the exact architecture, dynamics, and in some cases evolution history, of the planetary and planetesimal system.  This suggests that there is an opportunity to better constrain these properties of a system from the study of its exozodiacal dust.  In addition, \citet{bonsor2018} have shown that a radial break in the dust distribution (specifically for warm exozodiacal dust in that study) can indicate the presence of a planet and be used to constrain the planet's parameters.  \citet{rigley2020} have shown that the radial distribution and properties of (warm) exozodiacal dust can constrain the location of the source belt in case of supply through Poynting-Robertson drag.

Depending on the origin of the dust, it can also give critical insight into the environment in which rocky, potentially habitable planets exist.  For example, \citet{kral2018} and \citet{wyatt2020} have shown that the atmospheric composition of rocky planets can be strongly influenced by cometary impacts and that the outcome -- from barren planets to ocean worlds -- strongly depends on the impact energy and composition of the comets, i.e., their orbit, mass, and origin.  If comets are the source of exozodiacal dust, this may constitute an opportunity for better understanding the reasons behind the compositions of planetary atmospheres detected and characterized by future observatories such as HWO.  \citet{arras2022} have studied the accretion of dust onto planetary atmospheres which may also impact a planet's properties.


\section{Connection to exo-Earth imaging}
\label{sec:exo-earth}
Hot dust resides interior to the habitable zone, but could still impact the direct imaging of terrestrial planets within the habitable zone. Hot dust is likely delivered from further out in the planetary system.  We here specifically focus on a visible-light direct-imaging survey such as HWO.  Hence, the scattering properties of the dust grains at visible wavelengths are particularly important in this context.  We further consider a nulling-interferometric space mission observing at mIR wavelengths (e.g., the Large Interferometer For Exoplanets concept, LIFE, \citealt{quanz2022}).  The observability of exozodiacal dust with a focus on HZ dust and on observations with the Roman Coronagraphic Instrument was studied by \citet{douglas2019, douglas2022}.  The authors have shown that there is significant potential to detect and characterize dust in the outer HZs of a sample of stars, but that reaching closer in will be challenging.  We also note that there is a large body of research on the impact of HZ dust on exo-Earth imaging at visible wavelengths \citep[e.g.,][]{defrere2012b, roberge2012, stark2019, kammerer2022, morgan2024} and for a mIR-interferometric mission \citep[e.g.,][]{defrere2010, quanz2022}, but research on the impact of hot dust has so far been very limited.

There are then two possibilities for hot dust to impact such missions.  1) scattered light from material related to the hot dust moving through the HZ could present an additional background akin to more HZ dust, but could be particularly difficult to detect with mid-IR interferometry or  2) bright emission from the location of the hot dust (a few mas from the star) that introduces additional leakage around a coronagraph (visible, scattered light) or null leakage and confusion for a mIR interferometer (thermal emission).  These two cases are discussed in the following.

\subsection{Connection to habitable zone (HZ) and consequences}
As described in Sect.~\ref{sec:theory}, hot dust is likely comprised of small dust grains that may be inefficient emitters at mIR wavelengths, but efficient at scattering visible light.  The delivery mechanism for hot dust may then result in small, highly-scattering dust to be present in the habitable zone that is not easy to detect with mid-IR nulling interferometry (KIN, LBTI).  At first order, the thermal emissivity from dust grains is constant for wavelengths $\lambda < 2\pi a$, where $a$ is the grain size, and drops with the wavelength squared for $\lambda > 2\pi a$.  Hence, grains smaller than $1\,\mu$m do not emit efficiently at wavelengths longer than $\sim6\,\mu$m.

Assessing the presence and amount of small, highly scattering dust requires a better understanding of how hot dust is delivered into the inner system, but given the variability seen in some systems and the likely contribution from stochastic events such as comet disruptions/activity, it is likely there is dust moving through the habitable zone that could be clumpy and variable. More work is needed to quantify how much small, hot dust can be present without being detected in the mid-IR at the percent level, and what level of visible, scattered light from this dust could be expected.  This could be done based on existing photometric or interferometric limits in the mid-IR, new, more sensitive mid-IR spectroscopy, or by direct observations of the habitable zone and beyond with visible, scattered-light imaging with sufficient contrast, such as with the Roman CGI.  Since such dust would at most have a contribution to the thermal emission constrained by HOSTS in the same (general) wavelength range, the impact of this scenario on a LIFE-like mIR mission is reasonably well characterized through studies based on the HOSTS constraints \citep{quanz2022}.

\begin{figure}
  \centering
  \includegraphics[width=\textwidth]{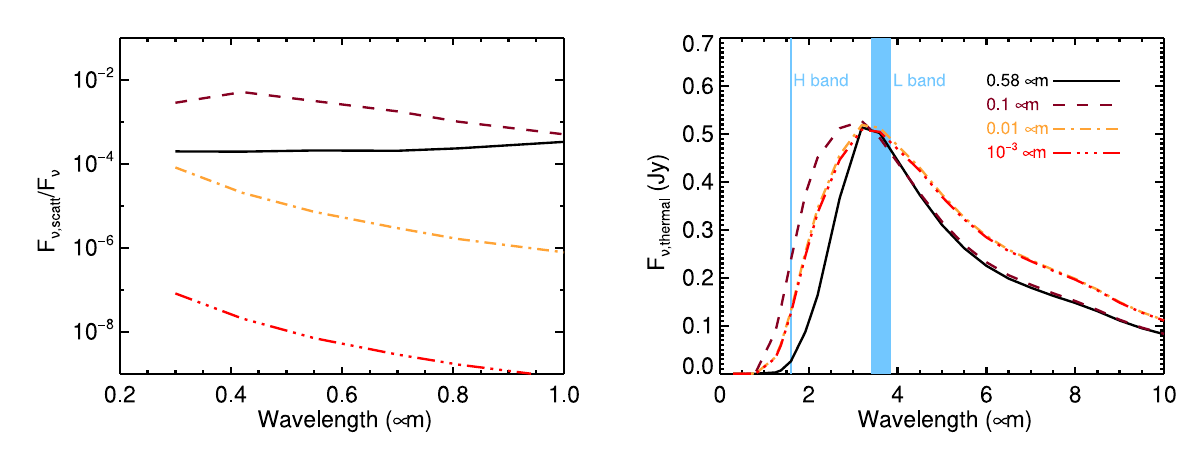}
  \caption{Scattered light predictions from the $\kappa$\,Tuc $H$-band and $L$-band observations. \emph{Left:} Scattered light emission ratio to the central star as a function of wavelength and grain size. Using one of the preferred models for the $\kappa$\,Tuc system \citep{kirchschlager2020}, we predict the integrated scattered light interior to $\sim$145\,mas as a ratio to the central star's flux density. The predicted emission varies by orders of magnitude.  \emph{Right:}  Plot of predicted mid-IR emission for different dust grain sizes. In all cases, we find that the predicted mid-IR emission matches that observed with MATISSE in the $L$~band, but varies differently across 2-10\,$\mu$m.  The scattered-light emission from the dust is to be compared to the total flux from the star and the anticipated suppression factor of the starlight for exo-Earth imaging (see Sect.~\ref{sect_coro_leakage} for details).}
  \label{fig_debes_ktucscatt}
\end{figure}

\subsection{Coronagraphic leakage \& Nulling Interferometric Leakage}
\label{sect_coro_leakage}
The impact from dust inside the inner working angle on precise PSF subtraction is already visible in current JWST observations of stars with circumstellar disks \citep{hinkley2023, boccaletti2024}.  For the specific case of hot exozodi, it can be assessed by estimating the scattered light contribution in the visible from hot dust in one of the few systems where there are some constraints on the location and properties of the dust, $\kappa$\,Tuc \citep{kirchschlager2020}. The worst case interpretation of the data corresponded to sub-micron amorphous carbon with a grain size of 0.58\,$\mu$m at temperatures of $\sim$900\,K and a mass of 7$\times$10$^{19}$\,g residing at 0.3\,au. For HWO, this corresponds to the scattered light component having an angular size of $\sim\lambda$/D. \citet{kirchschlager2020} noted that the grain size was not well constrained by the observations, so we have re-calculated their proposed model using MCFOST \citep{pinte06,pinte09} to predict the unresolved scattered light contribution from the dust as a function of grain size across a bandpass of 0.2-1\,$\mu$m, equivalent to the region where HWO might operate. $\kappa$\,Tuc was observed by \citet{marshall2016} using integral polarimetry, but constraints on individual targets from those observations are difficult due to the uncertain dust geometry (Sect.~\ref{sect_integral_polarimetry}).  Fig.~\ref{fig_debes_ktucscatt} shows the results for dust that ranges from 10$^{-3}$\,$\mu$m to 0.58\,$\mu$m. While the IR excess can be reproduced by a wide range of grain sizes, the scattered light is very sensitive to the typical grain size of the dust, and suggests that dust with radii of 0.1-1\,$\mu$m pose the largest threat to exoplanet imaging. In the worst case, coronagraph designs for HWO will need to suppress partially resolved light at $\lambda$/D by as much as 6 orders of magnitude at the inner working angle to remain robust against hot dust in the inner system of a star at the levels seen for $\kappa$\,Tuc. Leakage would also be an issue for 30\,m class ground-based NIR/Mid-IR high contrast imaging designs and for IR interferometry focused on the habitable zone (e.g., the LIFE concept).

$\kappa$\,Tuc's mid-IR emission, particularly between 2-10\,$\mu$m, is expected to be a few percent above the photosphere -- which is sufficient to be detected with high SNR spectroscopy using JWST/NIRSPEC or MIRI/MRS (I. Rebollido 2024, personal communication). Fig.~\ref{fig_debes_ktucscatt} demonstrates that simultaneous spectroscopic information between 2-10\,$\mu$m could better constrain the grain size of the dust, as the spectral shape of the thermal emission from different grains varies significantly. This could then place more stringent limits on the resulting scattered light. A mid-IR survey of HWO targets sensitive to $\sim$1\% excesses could help to characterize hot and warm dust better and provide reasonable predictions for the scattered light impact on direct imaging.

Hot exozodi is located at only a few stellar radii from the star.  This means that for nulling interferometry at 10\,$\mu$m the null depth will be affected due to the emission from the hot grains. An important contribution to the achievable null depth is the stellar leakage, which is increased by the presence of hot dust and its thermal emission.  Because the material is significantly farther from the stellar disk, and away from the bottom of the sinusoidal null fringe pattern ( $\phi \sim 2 \pi \sin{(\theta)} / \lambda $, where $\theta$ is the angle from the center of the star, and $\lambda$ is the wavelength),  the effect is much stronger for a given flux than the effect of a marginally resolved star. In order to obtain a specified null depth, either the emission from the hot exozodi must be small compared to the leakage from the star or the interferometric baseline must be shortened.  A simple calculation shows the hot zodi leakage should be less than $\sim$0.1\% of that from the star itself.  In addition the null measurement will be more sensitive to pathlength errors with the hot exozodi increasing the size of the resulting null fluctuations.  Further modeling and analysis of this issue is needed to clarify the importance of this effect to the noise budget for a nulling interferometer for exoplanet detection, like the TPFI/Darwin/LIFE mission concepts.  Furthermore, the issue may be further compounded by confusion added to the data from potentially clumpy dust structures (e.g., \citealt{defrere2010, defrere2012b, ertel2018SPIE}).

\section{Future observational prospects}
\label{sec:prospects}


\subsection{Structures and spectra of bright exozodi with MATISSE}  

New IR beam combiners on CHARA (MIRC-X, \citealt{anugu2020}, MYSTIC, \citealt{setterholm2023}) and VLTI (GRAVITY, \citealt{gravity2017}, MATISSE, \citealt{lopez2022}) do generally not reach the 1\% precision of PIONIER or FLUOR.  However, MATISSE is operating at longer wavelengths from the $L$~to $N$~band (where the $L$~band is the most relevant for exozodi since background-limited sensitivity becomes a concern at longer wavelengths with this beam combiner) and \citet{kirchschlager2020} have shown that an increasing dust-to-star flux ratio from the $H$~to $L$~band results in detectable excess at the precision MATISSE delivers.  This opens up the opportunity of MATISSE observations to survey stars for hot exozodi and characterize the new and previous detections.  MATISSE's spectral resolution could be used to search for spectral features and provides a denser u-v-coverage than previous observations, which increases the prospects of constraining the dust geometry, detecting structures in its distribution, and for the strongest detections potentially attempt image reconstruction.  Recent efforts with MATISSE have, however, shown that routinely obtaining high-S/N data from individual observations of exozodi is challenging (Ollmann et al., subm., Stuber et al., in prep., Priolet et al., in prep.), similar to previous observations in the nIR \citep{ertel2016}.  This illustrates that a major effort will be needed to fully exploit this capability.

\subsection{Luminosity function and variability at extreme sensitivity with NOTT} 

In the near future, Asgard/NOTT \citep{defrere2022, laugier2023, martinod2023} will become available at the VLTI.  NOTT is a new visitor instrument that will perform nulling interferometry in the $L$~band.  At an expected contrast performance at least at the level of the LBTI (order 0.1\%, \citealt{ertel2020SPIE}), this has the potential of being revolutionary for our understanding of hot exozodiacal dust and its connection to warm dust due to the intermediate observing wavelength and the sensitivity that will allow us to detect ten to fifty times\footnote{A factor ten comes from the improved high-contrast performance of nulling interferometry compared to constructive visibility measurements and a factor five comes from the improved dust-to-star flux ratio due to observations in $L$~band instead of the $H$~or $K$~bands.} more tenuous hot dust systems than currently possible and thus to derive a real luminosity function of the dust.  A strong involvement of exozodiacal dust experts in the project (e.g., work on observing strategy, pipeline optimization, target list, optimized instrument modes, etc.) has been enabled by an awarded NASA Astrophysics Decadal Survey Precursor Science grant (ADSPS, PI: S. Ertel, duration fall 2023 to fall 2026).  Improving the sensitivity of such thermal-infrared observations through advanced background-subtraction methods \citep{rousseau2024} may prove crucial for maximizing the science return from those observations.

\subsection{LBTI spectro-interferometry and nulling at L~and M~bands, all-sky L~and M-band survey with NOTT+LBTI}

The LBTI \citep{hinz2016, ertel2020SPIE} could be used for spectro-interferometry in Fizeau mode in the $L$~and $M$~band to detect hot exozodiacal dust, but this mode requires a major commissioning effort before it can be used for scientific observations.  This would provide the only currently available path towards exozodi observations in the $L$~or $M$~band in the Northern hemisphere.  A major detector upgrade of LBTI's Nulling-Optimized Mid-Infrared Camera (NOMIC, \citealt{hoffmann2014, ertel2022SPIE}) is currently planned (funding approved, PI:  K.~Wagner) and would enable an extension of the instrument's nulling capabilities towards the $L$~and $M$~bands after small, additional optics upgrades.  While LBT's baseline length is limited, observations at shorter wavelength and in Fizeau mode exploiting the full 22.7\,m edge-to-edge effective aperture allow for tracing hotter dust closer in than currently possible with $N$-band nulling.  This would enable a sensitive, nulling-interferometric all-sky survey for hot exozodiacal dust with both NOTT in the southern hemisphere and LBTI in the northern hemisphere.

\subsection{Aperture masking with JWST}
\label{sec:jwst_ami}
A new opportunity is presented by space-based aperture masking on JWST \citep{sivar2022} which is expected to deliver extremely high precision ($\sim0.1$\%) visibility measurements.  JWST's short baselines up to $\sim6\,$m, the large number of 21 visibility measurements taken simultaneously, and the wavelength coverage of 3-5\,$\mu$m may be able to detect the hot dust without over-resolving it and thus to put strong constraints on the geometry of its distribution.  A major challenge of JWST aperture-masking follow-up observations of detected hot exozodi is that bright stars observed and observable with OLBI quickly saturate on JWST (B. Pope, personal communication).

\subsection{Visible-light high-contrast imaging and imaging polarimetry}

Visible-light Extreme Adaptive Optics instruments on eight-meter class telescopes such as Spectro-Polarimetric High-contrast Exoplanet REsearch (SPHERE, \citealt{beuzit2019}) on the Very Large Telescope (VLT) or Subaru Coronagraphic Extreme Adaptive Optics (SCExAO, \citealt{Jovanovic2015}) Visible Aperture Masking Polarimetric Imager for Resolved Exoplanetary Structures (VAMPIRES, \citealt{norris2015}) on the Subaru telescope are equipped with polarimetric capabilities that could be used to both suppress the bright star light and spatially resolve and thus detect the polarimetric signal from the dust.  A 1\% excess at nIR wavelengths, if it stems entirely from gray-scattered light, would result in the same flux ratio at visible wavelengths.  Placing the dust at one resolution element ($\lambda/D$) from the star would dilute its signal over six resolution elements.  Optimistically assuming a polarization fraction of 30\%, this would then require a five-sigma contrast of $0.01/6\times0.3=5\times10^{-4}$ at 1\,$\lambda/D$ for a robust detection. Polarimetric observations with the Roman Space Telescope should also be explored \citep{anche2023}.  Roman coronagraphic observations would also have the capability to detect or rule out some of the most extreme scenarios related to scattered-light bright, mIR-faint dust moving through the habitable zone and of coronagraph leakage from hot exozodiacal dust, even if the performance is not expected to be sufficient to detect all cases that could critically impact HWO observations.  Finally, coronagraphic leakage from hot dust inside the inner working angle may be strong enough to be detected by Roman as a reduced contrast performance around hot-dust-hosting stars, which could provide an opportunity to test the severity of this issue.

\subsection{JWST and ground-based spectroscopy}

The recent results from precision JWST spectroscopy reported by \citet{worthen2024} using JWST/MIRI Medium Resolution Spectrograph observations of $\beta$\,Pic suggest that the strongest hot-dust excesses could be detected, confirmed, and characterized using this technique.  A characterization of the sensitivity, systematics, etc., of such observations may allow for independently confirming the hot-dust excesses using a different method than interferometry.  It may also allow for a new, efficient survey for hot exozodi and provide detailed spectral information (spectral slope, spectral features) of the dust that could constrain its properties (grain size, chemical composition).  Since many target stars are very bright (K magnitudes $\lesssim$5), similar ground-based observations using SpecX \citep{rayner2003} on the NASA Infrared Telescope Facility (IRTF) may be able to achieve similar sensitivity (C.~Lisse, personal communication), though a publication demonstrating this is still pending.  It is furthermore important to note that integral spectroscopy will likely not be able to well-distinguish between direct star light and star light scattered by dust grains and be thus unable to detect scattered light.  This provides an opportunity to better constrain the thermal emission from the dust alone and to constrain the amount of scattered light contributing to the nIR-interferometric detections in particular when obtaining (quasi-)simultaneous observations with both methods.

\section{Summary and conclusions}
\label{sec:summary}
We have reviewed the current knowledge in the field of hot exozodiacal dust.  We conclude that the phenomenon is most likely real and caused by hot circumstellar dust, but point out that conclusive proof in particular of the latter statement is still missing.

Near-infrared optical interferometry has so far been the only method to reliably detect the excess emission, though the potential of variability of the excess and the fact that the signals are often detected close to the sensitivity limit where random errors may move the excess in a single observation above or below the detection threshold make repeated detections challenging.  Currently, VLTI/MATISSE is producing the most impactful new observations with this method.  The most promising  new methods to detect and further characterize hot exozodi are  (1) precision mid-infrared spectroscopy with JWST,  (2) $L$-band nulling interferometry with VLTI/NOTT and an upgraded LBTI.  These observations have the potential to conclusively determine the veracity and nature of the excess detections and should be pursued at high priority.  Coronagraphic observations with the Roman Space Telescope could rule out some of the worst-case scenarios of the hot-exozodiacal-dust impact on future exo-Earth imaging missions.

We also conclude that there exists as of yet no theoretical explanation for the presence of the hot dust.  In particular, current scenarios all produce too much mid-infrared emission compared to existing observational constraints.  This is typically caused by the fact that the material must move through cooler regions of the system either during delivery or at least during removal from the system by radiation pressure blow-out, and during that time the material would emit significantly at mid-infrared wavelengths.  Given the lack of an adequate understanding of the dust origin, we are also unable to predict the occurrence rate of the hot dust at levels below the current detection limits (luminosity function).  We are also unable to constrain how much dust could be expected to move through the habitable zone of a system that hosts a hot exozodi during dust delivery and removal.  Depending on the dust properties, such dust might scatter visible star light very efficiently but not be detectable by mid-infrared nulling interferometry.

Using a toy-model prediction of the scattered light from hot exozodiacal dust, we find that hot dust puts strong requirements on the performance of a coronagraphic exo-Earth imaging mission.  In a worst-case scenario, scattered light from hot dust located at the coronagraph's inner working angle would have to be suppressed by as much as six orders of magnitude.  Similarly, the hot dust is expected to strongly affect a mid-infrared interferometric space mission to detect and characterize potentially habitable exoplanets.  While there are less fatal scenarios, it is at least plausible and not unlikely that the presence of hot exozodiacal dust in a system will critically limit the sensitivity of exo-Earth imaging observations around that star.  A better understanding of the phenomenon is thus crucial for the success of any such mission and should be pursued with high priority.  At the same time, we find that exozodiacal dust has the potential to provide critical context for the detection of potentially habitable exoplanets as it can constrain the architecture, dynamics, and evolution history of planetary systems and hence shed light onto the environment in which such planets exists.

\section{Acknowledgements}
\label{sec:acknowledgements}
This work is supported by NASA through grants 80NSSC21K0394 (PI: Ertel, supports S.E., V.C.F., H.R.), 80NSSC23K1473 (PI: Ertel, supports S.E., V.C.F., W.C.D., H.R., J.S., T.A.S.), and 80NSSC23K0288 (PI: Faramaz, supports V.C.F., T.A.S., S.E.).
~
T.D.P.\ is supported by a UKRI Stephen Hawking Fellowship and a Warwick Prize Fellowship, the latter made possible by a generous philanthropic donation.
~
D.D.\ has received funding from the European Research Council (ERC) under the European Union's Horizon 2020 research and innovation program (grant agreement CoG - 866070)
~
Y.H.\ was supported by the Jet Propulsion Laboratory, California Institute of Technology, under a contract with the National Aeronautics and Space Administration (80NM0018D0004).
~
F.K.\ has received funding from the European Research Council (ERC) under the European Union’s Horizon 2020 research and innovation programme DustOrigin (ERC-2019-StG-851622).
~
This work is supported by the NASA Exoplanet Exploration Program (ExEP) and the Exoplanet Program Analysis Group (ExoPAG) as part of the ExoPAG Study Analysis Group 23 (SAG23).
~
This research has made use of the Astrophysics Data System, funded by NASA under Cooperative Agreement 80NSSC21M00561.  We have made extensive use of the matplotlib Python package \citep{hunter2007}.

\bibliographystyle{aasjournal}
\bibliography{biblio}

\end{document}